\documentclass[prl, twocolumn, a4paper, superscriptaddress, longbibliography]{revtex4-2}

\usepackage{amsmath}
\usepackage{amssymb}
\usepackage{amsthm}
\usepackage{graphicx}
\usepackage{enumitem}   
\usepackage[font={small, sf}, format=plain, justification=justified, labelfont=bf]{caption}
\usepackage{float}

\usepackage{nicematrix}

\usepackage{xcolor}
\usepackage{hyperref}
\usepackage{tikz}
\usepackage{mhchem}
\usepackage{blkarray}

\usepackage{tabularray}
\usepackage{layouts}

\usepackage{easyReview}

\DeclareMathAlphabet{\mathsfit}{T1}{\sfdefault}{\mddefault}{\sldefault}
\SetMathAlphabet{\mathsfit}{bold}{T1}{\sfdefault}{\bfdefault}{\sldefault}

\renewcommand{\S}{\mathbb{S}}
\newcommand{\M}{\mathbb{M}}

\newcommand{\affrho}{\mathcal{A}_{\rho}}
\newcommand{\dQ}{\mathrm{d}_Q}

\renewcommand{\d}{\mathrm{d}}

\renewcommand{\ss}{\text{\,}}
\newcommand{\gX}{\mathbf{g}_X}
\newcommand{\Js}{\mathcal{J}}
\newcommand{\Z}{\mathcal{Z}}
\renewcommand{\log}{\ln}

\newcommand{\Sex}{\boldsymbol{\nabla}^\text{ex}}
\newcommand{\hatgX}{\widehat{\boldsymbol{g}}_\textsf{X}}
\newcommand{\eA}[1]{e^{\mathcal{A}_{#1}^\ast}}

\newcommand{\C}{\mathcal{C}}
\renewcommand{\P}{\mathcal{P}}
\renewcommand{\L}{\mathcal{L}}
\newcommand{\E}{\mathcal{E}}
\newcommand{\R}{\mathcal{R}}
\newcommand{\Eex}{\mathcal{E}^\text{ex}}
\newcommand{\Pex}{\mathcal{P}^\text{ex}}

\DeclareTextFontCommand{\sf}{\textsf}
\DeclareTextFontCommand{\quest}{\bfseries\color{blue}}

\begin{document}
	
	\title{
		Structural constraints limit the regime of optimal flux in autocatalytic reaction networks
	}

	\author{Armand Despons}
	\affiliation{Gulliver Laboratory, UMR CNRS 7083, PSL Research University, ESPCI, Paris F-75231, France}

	\author{Yannick De Decker} 
	\affiliation{Center for Nonlinear Phenomena and Complex Systems (CENOLI), Université libre de Bruxelles (ULB), Campus Plaine, C.P.
		231, B-1050 Brussels, Belgium}
	
	\author{David Lacoste}
	\affiliation{Gulliver Laboratory, UMR CNRS 7083, PSL Research University, ESPCI, Paris F-75231, France}

	\date{\today}

	\maketitle
	
	\section*{Abstract}
	
	Autocatalytic chemical networks play a predominant role in a large number of natural systems such as in metabolic pathways and in ecological networks. 
	Despite recent efforts, the precise impact of thermodynamic constraints on these networks remains elusive.
	In this work, we present a theoretical framework that allows determining bounds on the thermodynamic affinity and on the concentrations of autocatalysts in mass-action autocatalytic networks. 
	These bounds can be obtained solely from the stoichiometry of the underlying chemical reaction network, and are independent from the numerical values of kinetic parameters. 
	This property holds in the specific regime where all the fluxes of the network are tightly coupled and maximal. 
	Our method is applicable to large networks, and can be used to complement constraints-based modeling methods of metabolic networks, which typically do not provide predictions about thermodynamic properties or concentration ranges of metabolites.

	\section*{Introduction}
	
	The dynamical properties of most systems found in nature can be traced back to a combination of chemical reactions that are maintained out of equilibrium. This is particularly the case in biology and in ecology, where these chemical reaction networks can reach high levels of complexity. Many such networks involve autocatalysis, which is the ability of chemicals to catalyze their own formation \cite{Schuster2019,Xavier2020}. Autocatalysis enables exponential growth \cite{Lin2020,roy2021}, self-replication \cite{Ameta2021}, and metabolism \cite{Lancet2018}. 
	
	The properties of chemical reaction networks are constrained by non-equilibrium thermodynamics. For example, living systems are thought to self-organize thanks to energy and matter flows, which allow them to lower their own entropy at the expense of an entropy increase in the environment \cite{Schrodinger1944}, as required by the second law. Thermodynamics is also believed to play a key role in chemical evolution \cite{Kolchinsky2021, Pascal2013} and in the organization of ecological communities \cite{Endres2017, George2023}. Yet, understanding precisely the role played by thermodynamic constraints proved difficult both for biology and ecology, despite the many recent efforts on this issue \cite{Cengio2022, Avanzini2021, Hirono2021, Sughiyama2021, Ito2021}. Because of this, we still do not fully understand the fundamental principles by which chemical evolution operates, which limits our ability to design new materials with life-like properties.
	
	This lack of general understanding also has more practical consequences. The knowledge we have about metabolic networks is rather limited as far as kinetics is concerned. For this reason, current modeling approaches for metabolic networks focus on predicting steady fluxes by optimizing an objective function with linear constraints, as in Flux Balance Analysis  \cite{Orth2010, Fell1997} or in Structural Kinetic Modeling  \cite{Steuer2006}. These methods are general, valid for any chemical network and do not require a detailed knowledge of the kinetics, but it is not easy to use them to make predictions about metabolite concentrations. For instance, at the moment, concentration ranges of metabolites can be predicted from these methods but only for mass-action networks \cite{kuken2019}. 
	
	The aforementioned problems are all related to the fact that while non-equilibrium thermodynamics is a well-established discipline, its implications for autocatalytic networks have not been fully explored yet. In this context, we develop a theoretical framework for autocatalytic chemical networks operating in a stationary non-equilibrium regime. This approach builds on a recent stoichiometric classification of autocatalytic chemical networks \cite{Blokhuis2020, Unterberger2022}, which can be used to identify such networks thanks to chemoinformatic techniques \cite{andersen2021, Peng2022, Arya2022}. We find that in a specific regime where fluxes in the network are tightly coupled and optimal, there is a connection between the topology of a network, the stoichiometry of its autocatalytic reactions, and the thermodynamic force keeping this network out of equilibrium. In this regime, we show that the force required to operate an autocatalytic network at a maximum rate obeys universal constraints, which depend on topology and stoichiometry, but are independent from the kinetic rate constants of the reactions. Because these constraints contain information on the topology of the network, they can be used to rule out certain network architectures for a given global autocatalytic reaction, even in the absence of any knowledge about kinetics. 
	In the end, our work shows how thermodynamics constrains the chemical space accessible to autocatalysis, which is relevant for chemical evolution and Origin of Life studies. 
	It also provides relevant information regarding the thermodynamics of autocatalytic networks and about the concentration ranges of the autocatalysts, which could be useful to design new chemical networks.

	\section*{Results} 
	
	\subsection*{Motivating example}
	
	
	Our objective is to find a connection between the rate of production of an autocatalytic process, its distance from equilibrium, and the stoichiometry of the underlying reaction network. 
	Consider, for illustration, the following reactive system:
	
	\NiceMatrixOptions{
		code-for-last-row = \color{black!70}\scriptstyle,
		code-for-first-col = \color{black!70}\scriptstyle,
	}
	
	\begin{equation}
		\begin{split}
			\label{eq:type_I_network}
			\ce{\textsf{F} + $\alpha$ \, \textsf{A}& <-->[$k_{+1}$][$k_{-1}$]   \textsf{B}}\\
			\ce{\textsf{B} & <-->[$k_{+2}$][$k_{-2}$] $\beta$\, \textsf{A} + \textsf{W}, }
		\end{split} 
	\end{equation}
	with $\beta > \alpha >0$.
	The topology of \eqref{eq:type_I_network} is encoded in its stoichiometric matrix,
	\begin{align}
		\label{eq:full_stoich_Type_I}
		\nabla = ~ 
		\begin{pNiceMatrix}[last-row, first-col]
			\textsf{F} & -1 & 0 \\
			\textsf{W} & 0 & 1  \\
			\textsf{A} & -\alpha & \beta \\
			\textsf{B} & 1 & -1 \\
			& \mathsfit{1} & \mathsfit{2} \\
		\end{pNiceMatrix}. 
	\end{align}
	The species $\sf{F}$ and $\sf{W}$ act as fuel and waste for the overall production of the other species. In what follows, we will treat their concentrations, $f$ and $w$, as constants. 
	Only a submatrix of \eqref{eq:full_stoich_Type_I} is required to capture the autocatalytic behavior of \eqref{eq:type_I_network}:  
	\begin{equation}
		\label{eq:stoich_Type_I}
		\mathbb{S} = ~ 
		\begin{pNiceMatrix}[last-row, first-col]
			\textsf{A} & -\alpha & \beta \\[0.5em] 
			\textsf{B} & 1 & -1 \\
			& \mathsfit{1} & \mathsfit{2} \\
		\end{pNiceMatrix}.
	\end{equation}
	This submatrix establishes a connection between the time derivative of the concentrations of the autocatalytic species $a$ and $b$ with the reaction fluxes,
	\begin{equation}
		\label{eq:dyn_type_I}
		\d_t \, \begin{pmatrix}
			a \\[0.5em] b
		\end{pmatrix} = \mathbb{S} \cdot \mathbf{j},
	\end{equation}
	where  
	$\mathbf{j}$ is a vector containing the fluxes of reactions 1 and 2,  $\mathbf{j}=(j_1, j_2)^\intercal$.
	For ideal isothermal solutions, each of these reaction fluxes can be decomposed as the difference of two one-way fluxes obeying the law of mass action: 
	\begin{align}
		& j_{+1} = k_{+1} \, f \, a^\alpha, & 
		j_{-1} = k_{-1} \, b, \\
		& j_{+2} = k_{+2} \, b, &    j_{-2} = k_{-2} \, a^\beta\,  w ;
	\end{align}
	and the components of $\mathbf{j}$ are $j_{i} = j_{+i} - j_{-i}$.
	Nothing prevents this reactive system from reaching the equilibrium state, since the concentrations of both $\sf{A}$ and $\sf{B}$ are unconstrained.  
	To drive the network \eqref{eq:type_I_network} away from its equilibrium state, we 
	introduce a control mechanism which maintains the concentrations $a$ or $b$ constant 
	thanks to an outgoing flux. 
	
	\begin{figure*}
		\centering
		\includegraphics{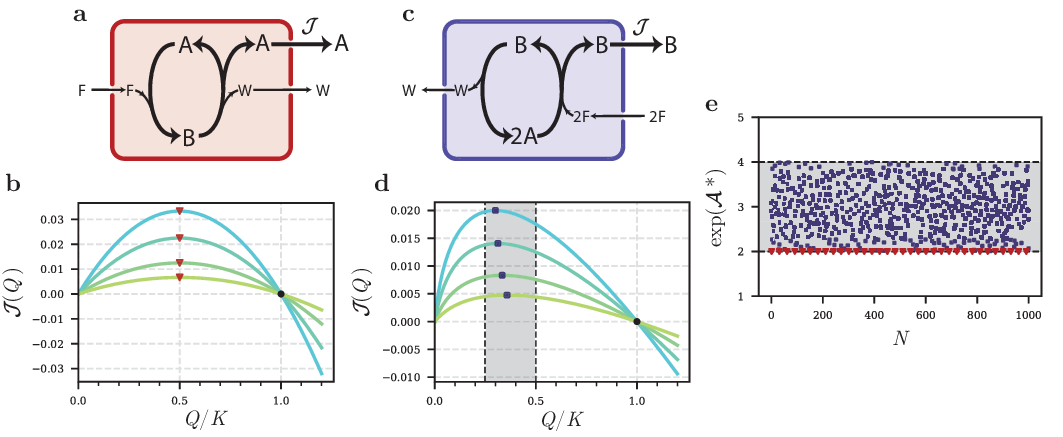}
		\caption{\label{fig:Type_I_exemple}  
			Selected properties of the autocatalytic system \eqref{eq:type_I_network} for $\alpha=1$ and $\beta= 2$. 
			\textbf{a} Representation of a reactor in which the concentrations of $\sf{F}$, $\sf{W}$ and $\sf{A}$ are maintained constant thanks to exchanges with external reservoirs (chemostats).   
			\textbf{b} The corresponding global flux $\Js$ as a function of $Q$ vanishes when $Q = K$ (on the black circle) and reaches a maximum at the same location $Q^\ast = K \slash 2$  (red triangles). 
			\textbf{c} Species $\sf{F}$, $\sf{W}$ are still chemostatted, but now $\sf{B}$ is chemostatted instead of $\sf{A}$. 
			\textbf{d} Now the maxima of the corresponding global reaction rate are reached at different locations such that $ 1 / 4 \leq Q^\ast/K \leq  1 / 2$ (shaded grey area). Note that all the fluxes still vanish when $Q = K$. 
			\textbf{e} Cloud of points for the exponential of the global affinity $\mathcal{A}^*$, for randomly chosen sets of kinetic rate constants indexed by $N$.  
			When species $\sf{A}$ is controlled (red triangles), the optimal affinity is always given $\ln 2$, as explained in the main text. 
			When species $\sf{B}$ is controlled (blue squares), the optimal affinity admits both a lower and upper bound, which corresponds to the grey area in the lower plot of \textbf{b}. 			
		}
	\end{figure*}
	
	
	When the concentration of species $\sf{A}$ is controlled (chemostatted), the dynamics of the system is entirely controlled by the evolution law for $b$, $\d_t\,  b   = j_1 - j_2$. 
	The steady state is  
	such that $j_1^\ss = j_2^\ss$, which shows that 
	there is a tight coupling \cite{Wachtel2018} between the two reactions. Here, $j_1^\ss$ and $j_2^\ss$ are also equal to the production rate $\mathcal{J}^\ss$ of the overall reaction:  
	\begin{equation}
		\label{eq:ss_reaction_Type_I_A}
		\sf{F} + \alpha\,  \sf{A}  \ce{ ->} \beta\, \textsf{A} + \sf{W }.
	\end{equation} 
	This reaction has an overall affinity (which is equal to the opposite of the Gibbs free energy $\Delta G$) $\mathcal{A}= \mu_\sf{F}  - \mu_\sf{W}- (\beta-\alpha)\,  \mu_\sf{A}$, where $\mu_i$ is the chemical potential of species $i$. Since $\mu_\sf{F}$ and $\mu_\sf{W}$ are fixed, fixing $\mu_\sf{A}$ is essential to maintain the system in a non-equilibrium state where $\mathcal{A}$ is non-zero. 
	Otherwise, the system will reach equilibrium, where the affinity vanishes.  
	Solving for the steady-state concentration provides the expression for the macroscopic flux of production of species $\textsf{A}$: 
	\begin{equation}
		\mathcal{J}^\ss  = \dfrac{k_{+1} \, k_{+2}}{k_{-1} + k_{+2}}  \left[ a^\alpha\,  f - \left(\dfrac{k_{-1} k_{-2}}{k_{+1} k_{+2}}\right) ~ a^\beta \, w \right].
	\end{equation}
	Introducing the reaction quotient $Q = a^{(\beta-\alpha)}\,  w / f$ and the equilibrium constant $K = k_{+1} k_{+2} / k_{-1} k_{-2}$ of the global reaction \eqref{eq:ss_reaction_Type_I_A}, it is easy to show that the global flux $\Js$ has the same sign as the affinity, since  $\mathcal{A} = \log \left( K/Q \right) $ (we work with units where $R\, T=1$).
	Furthermore, the current goes through a maximum as a function of $a$ whenever 
	\begin{equation}
		\dfrac{a^{(\beta-\alpha)} \, w}{f} = \dfrac{\alpha}{\beta} \, \dfrac{k_{+1}\,  k_{+2}}{k_{-1} \, k_{-2}},
	\end{equation}
	which corresponds to the condition $Q=Q^* = \alpha\slash \beta \times K$. 
	Thus, the maximum rate is reached when the chemical affinity becomes
	\begin{equation}
		\label{eq:chem_aff_Type_I}
		\mathcal{A^*} =  \ln \frac{\beta}{\alpha}.
	\end{equation}
	Hence, the distance from equilibrium at which the autocatalytic network achieves its optimal production rate, namely $\mathcal{A^*}$, is not fixed by the values of kinetic constants or by the equilibrium constant of the global process. It only depends on the stoichiometry of the overall reaction.  We illustrate these results in Fig. \ref{fig:Type_I_exemple}, for the case $\alpha=1, \beta=2$. 
	Note that this condition on the affinity can also be expressed in terms of the concentration $a$. Since $\exp (\mathcal{A}^*)=K / Q^*$, the point of maximum flux is given by 
	\begin{equation}
		\left( \frac{a^*}{a_{\text{eq}}} \right)^{\beta - \alpha}=  \frac{\alpha}{\beta},
	\end{equation}
	which means that optimality is reached when the concentration of the chemostat is at half its equilibrium value.

	
	We can carry a similar analysis if, instead of \textsf{A, B} is the chemostatted autocatalytic species. Now, the dynamics is ruled by 
	$d_t \, a  = \beta \, j_2 - \alpha \, j_1$, and the steady state is such that $j_1^\ss \slash \beta= j_2^\ss \slash \alpha =\mathcal{J}^\ss$, which  is the overall production rate of species \textsf{B}, 
	\begin{equation}
		\label{eq:ss_reaction_Type_I_B}
		\beta\, \textsf{F} + \alpha\, \sf{B} \ce{->} \beta\, \textsf{B} + \alpha\, \sf{W}.
	\end{equation}
	The steady-state solution now involves polynomials of different order, making it impossible to find explicitly the conditions maximizing the rate of production with the previous method. Nonetheless, we can determine $\mathcal{A}^*$ by numerically finding the value of $Q^\ast$ that maximizes the global reaction rate for randomly generated values of the various kinetic constants. We find that the optimal affinity is now bounded from below  (see Supplementary Note 12): 
	\begin{equation}
		\label{eq:ex_chemB}
		\mathcal{A}^\ast \geq \ln \frac{\beta}{\alpha}.
	\end{equation}
	Here too, the constraint acting on the optimal distance from equilibrium solely depends on the stoichiometry of the overall reaction. 
	This thermodynamic constraint translates into a threshold for the value of the concentration of the controlled species, $b^\ast$ as
	\begin{equation}
		\label{eq:ex_chemB-equiv}
		\left( \frac{b^\ast}{b_{\text{eq}}} \right)^{\beta-\alpha} \leq \frac{\alpha}{\beta},
	\end{equation}
	where $b_{\text{eq}}$ is the equilibrium concentration of $\textsf{B}$. Note the importance of the condition $\beta > \alpha>0$, which guarantees the invertibility of the stoichiometric matrix $\mathbb{S}$ and leads to 
	equalities or inequalities Eqs.~\eqref{eq:chem_aff_Type_I}-\eqref{eq:ex_chemB-equiv}.
	
	Figure \ref{fig:Type_I_exemple} summarizes the behavior of the network Eq.~\eqref{eq:type_I_network} in the particular case $\alpha = 1$ and $\beta = 2$. 
	It also reveals that, on top of the upper bound Eq.~\eqref{eq:ex_chemB-equiv}, the location of the global flux is also lower-bounded, $Q^\ast / K \geq 1/4$, yielding an upper bound for the affinity $\mathcal{A}^\ast$. 
	This behavior can be recovered analytically for arbitrary values of $\alpha$ and $\beta$ (see Supplementary Notes 11 \& 12).

	\subsection*{General approach for autocatalytic chemical reaction networks (CRNs)}
	
	In what follows, we present a general approach that enables the computation of the overall affinity $\mathcal{A}^*$ corresponding to a condition of local extremum of the global production rate. Our method relies on the stoichiometry of the reaction network, and circumvents the need to explicitly evaluate the steady states or reaction rates of the system under consideration. 	
	

	Following a recent stoichiometric characterization of autocatalysis \cite{Blokhuis2020}, a general autocatalytic network should contain one or several autocatalytic cores. 
	An essential feature of these cores is that they are described by an invertible stoichiometric sub-matrix.
	The existence of this invertible sub-matrix is a sufficient condition for autocatalysis, which implies the absence of conservation laws \cite{Avanzini2021,Rao2016, Polettini2014} thanks to Gordan's theorem \cite{Blokhuis2020}. In the following, we assume that the full stoichiometric matrix $\boldsymbol{\nabla}$ contains a square submatrix $\S$ that is invertible~: 
	\begin{equation}
		\label{eq:Nabla}
		\boldsymbol{\nabla} = \quad 
		\begin{pNiceArray}{w{c}{1em}c:w{c}{1em}c}
			\Block{2-4}{\Sex} &  &  & \\
			& & & \\
			\hdottedline
			\Block{2-2}{\S} &  & \Block{2-2}{ \mathbf{0} } &   \\
			\vspace*{0.5em}
		\end{pNiceArray}
	\end{equation}
	Reactions of $\S$ will be called autocatalytic reactions, species of $\S$ will be called autocatalytic species, while other species will be called external. 
	Note that we do not require that the matrix $\S$ corresponds necessarily to one of the autocatalytic cores listed in Table \ref{tab:Recap_core} of the Methods Section.  
	Further, we also allow the network to contain catalytic reactions, (\emph{i.e.} reactions where the same species can be both reactants and products) in contrast with the assumptions of \cite{Blokhuis2020}.
	The presence of a block matrix of zero in the lower right part of the matrix $\boldsymbol{\nabla}$ in Eq.~\eqref{eq:Nabla} means that there is no boundary flux to or from external species. 
	Such a splitting of species into internal and external ones is a standard assumption in metabolic network analysis \cite{Qian2005}.

	Similar to the example above, the concentrations of all external species are assumed to be constant, either because these species are in excess, which is typically the case for food or fuel species, or because they are in contact with a reservoir (chemostatted). 
	Given the structure of the full stoichiometric matrix of Eq.~\eqref{SM-eq:Nabla}, one can show that the network can still be in detailed balance (see Eq.~\eqref{SM-eq:Balance_chemical_potential} and Supplementary Note 1). 
	To break detailed balance, in addition, the concentration of at least one of the autocatalytic species should be controlled, by actively maintaining its concentration constant with an outgoing flux.

	We call this special species the $X$ species, and we call the remaining non-controlled autocatalytic ones the $Y$ species. 
	The stoichiometric matrix $\mathbb{S}$ splits into a row vector ${\bold S}^{X}$ and a matrix $\mathbb{S}^{Y}$ \cite{Rao2016}: 
	\begin{equation}	
		\label{eq:split_stoich_matrix}
		\mathbb{S} = \begin{pmatrix}  {\bold S}^{X} \\  \mathbb{S}^{Y} \end{pmatrix}. 
	\end{equation}
	The corresponding kinetic equations are given by 
	\begin{align}
		\d_t \, x  & = {\bold S}^X \cdot \mathbf{j} + I = 0 \label{eq:dyn_internal_CRN_1} \\
		\d_t \, \mathbf{y} &  = \S^Y \cdot \mathbf{j} 	\label{eq:dyn_internal_CRN_2},
	\end{align}
	where $x$ denotes the concentration of species $X$ and $\mathbf{y}$ is the concentration vector of all the $Y$ species. 
	The vector $\mathbf{j}$ contains the rates of the autocatalytic reactions, and $I$ is a scalar function describing the exchange of matter with the chemostat. 
	
	The inverse $\S^{-1}$ plays an important role: 
	\begin{equation}
		\S^{-1} = \left\lbrace  \mathbf{g}_{\sigma} \right\rbrace_{Z_\sigma \in  \Z },
	\end{equation} 
	where $\mathbf{g}_\sigma $ is the column of $\mathbb{S}^{-1}$ associated to species $Z_\sigma$, which denotes autocatalytic species of the $X$ or of the $Y$ type, and $ \Z $ is the set of all the autocatalytic species.
	It follows from the property of the inverse that $\mathbf{g}_\sigma $ represents a reaction pathway that produces a single unit of species $Z_\sigma$ without affecting the other species (see Supplementary Note 2). 
	For this reason, $\mathbf{g}_\sigma$ is the elementary mode of production of species $Z_\sigma$ \cite{Blokhuis2020}. 
	From all the elementary modes of production, one can build a reaction vector $ \mathbf{g} = \sum_\sigma \mathbf{g}_\sigma$, which represents a combination of elementary modes that increases the amount of all the autocatalytic species by one unit: 
	\begin{equation}
		\label{eq:def_g}
		\S \cdot \mathbf{g} = \mathbf{1},
	\end{equation}
	where $\mathbf{1}$ is a column vector full of ones. 
	The existence of this vector is a sufficient condition for autocatalysis. 
	
	The steady-state of Eq.~\eqref{eq:dyn_internal_CRN_2} is a vector belonging to the right nullspace of $\S^{Y}$ \cite{Rao2016, Polettini2014}, which is spanned by $\gX$, the elementary mode of production associated to species $X$. 
	Thus, the stationary flux vector can be written as 
	\begin{equation}
		\label{eq:steady_current}
		\mathbf{j}^\ss = \Js ~ \gX,
	\end{equation}
	which we call a tight coupling condition, since all elementary fluxes are proportional to the global production rate $\mathcal{J}$ with constant coefficients of proportionality.
	Since in general, $\gX$ has rational components, it is convenient to rescale this vector to facilitate its interpretation as a mode of production. 
	We show in the Methods section that, after the rescaling, the bounds on the global affinity are also rescaled by the same amount.
	
	
	We now introduce assumptions about the dynamics of the network at the level of elementary reactions. Every such reaction denoted $\rho$ is assumed to be reversible, with a net flux given by  
	\begin{equation}
		j_\rho = j_{+\rho} - j_{-\rho}.
	\end{equation}
	Here, one-way fluxes $ j_{\pm \rho}$ obey mass-action kinetics  \cite{Feinberg2019, Pekar2005}:
	\begin{equation}
		\label{eq:mass_action_kin}
		j_{\pm \rho} = k_{\pm \rho} \, \prod_\sigma \, z_\sigma^{S_{\pm \rho}^\sigma},
	\end{equation}
	where $z_\sigma$ is the concentration of species $Z_\sigma$, and $ \S_+ $ (resp. $\S_-$) is the stoichiometric matrix associated to forward (resp. reverse) reactions, such that $\S = \S_- - \S_+$.
	Note that the constant concentrations of the fuel/food species have been absorbed in the effective rate constants $k_{\pm \rho}$. We choose to work with ideal solutions for clarity of presentation, but we show in the Supplementary Note 4 that the results presented below remain valid for non-ideal solutions. 
	
	\subsection*{Overall affinity}
	
	The affinity of an elementary reaction is connected to fluxes by means of the flux-force relationship: ${\mathcal{A}_\rho=\log{\left(j_{+ \rho} \slash j_{- \rho} \right)}}$. 
	Taking the linear combination of the elementary affinities, we obtain the global affinity 
	\begin{equation}
		\label{eq:overall_affinity_main}
		\mathcal{A}= \sum_\rho g_X^\rho \, \mathcal{A}_\rho = \sum_\rho g_X^\rho \, \log{\left(\frac{j_{+ \rho} }{j_{- \rho} }\right)} = \log \left( \dfrac{K}{Q} \right),
	\end{equation}
	where $K$ and $Q$ are respectively the equilibrium constant and the reaction quotient $Q$ \cite{Prigogine2014}.
	Here, the reaction quotient coincides with the concentration of the controlled autocatalytic species ($Q=x$) because the concentrations of food species have been incorporated in the rate constants, and the stationary global flux is a function of this quantity, $\Js=\Js(Q)$. 
	
	We expect that in the cases of interest, this flux will present at least one maximum when the CRN is brought out of equilibrium. Indeed, due to the tight coupling condition, the total entropy production rate (EPR) at steady-state has a simple expression \cite{Polettini2014, Rao2016, Avanzini2021}: 
	\begin{equation}
		\label{eq:EPR_main}
		\Sigma = \sum_\rho  j_\rho\, \mathcal{A}_\rho = \mathcal{J} \mathcal{A}.
	\end{equation}
	The second law of thermodynamics dictates that $\Sigma \geq 0 $, and equality is achieved at equilibrium, where $Q=K$ and both $\mathcal{J}$ and $\mathcal{A}$ vanish. 
	It follows, from Eq.~\eqref{eq:overall_affinity_main}, that $\Js(Q) \geq 0 $ when $ Q \in [0, K] $.
	Let us consider a class of autocatalytic networks, 
	for which species $X$ has a nucleating role \cite{Peng2022}. In that case, the flux $\mathcal{J}( 0 )$ is zero when this species is absent (in other words, for $Q = 0$). 
	This was the case, for example, with the simple autocatalytic system discussed earlier (see Figure \ref{fig:Type_I_exemple}). For these networks, there must be at least one local extremum of the global flux in the interval $Q \in [0, K]$. 
	This regime is optimal, in the sense that species $X$ is produced at a maximal rate. 
	Importantly, the extremum of the global flux and that of the EPR do not coincide because $\mathcal{A}$ in Eq.~\eqref{eq:overall_affinity_main} is a non-linear function of $Q$ which also enters in Eq.~\eqref{eq:EPR_main}. 
	This is in agreement with the observation that the optimal regime of operation of non-equilibrium systems generally does not correspond to a point where the EPR is extremum \cite{Baiesi2018}. 
	
	\subsection*{Response coefficients at maximum flux}
	
	We call  $Q^\ast$ the value of $Q$ which makes the global flux maximal, with a zero derivative: $\d_Q\, \Js  = 0$. 
	Due to the tight coupling condition Eq.~\eqref{eq:steady_current}, all reaction fluxes are also at an extremum at $Q^\ast$: ${\d_Q \, j^\ss_\rho  = {\d_Q \, j^\ss_{+\rho}}-{\d_Q \, j^\ss_{-\rho}} = 0}$, for any reaction $\rho$.  
	To characterize this configuration, we introduce the log-derivative of the stationary elementary fluxes, which we call $F_{\pm \rho}$ :
	\begin{equation}
		\label{eq:definition_F}
		F_{\pm \rho} = \dQ \log j_{\pm \rho}^\ss = \sum_\sigma S_{\pm \rho}^\sigma  \, \dQ \log z_\sigma^\ss.
	\end{equation}

	\begin{figure}[h]
		\centering	
		\includegraphics{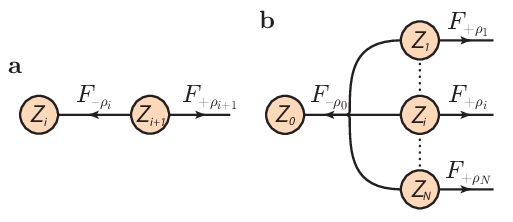}	
		\caption{\label{fig:Kirchhoff} 
			Structural constraints acting on the log-derivatives of the steady unidirectional fluxes with mass-action kinetics.
			\textbf{a} 
			When a species $Z_{i+1}$ is connected by unimolecular pathways upstream and downstream, the two outgoing one-way fluxes  have equal log-derivative, $F_{- \rho_i} = F_{+\rho_{i+1}}$.
			\textbf{b} 
			In a branched pathway, a species $Z_0$ is connected to various species $Z_i$, and the sum of the log-derivative of the forward flux of all products balance the log-derivative of the reverse branched reaction, $F_{-\rho_0} = \sum_i S_{-\rho_0}^i \, F_{+\rho_i}$.
		}
	\end{figure}
	
	These are the response coefficients of the steady unidirectional fluxes with respect to a change in $Q$. 
	Because the $F_{\pm\rho}$s are log-derivatives, all the factors entering rate laws that do not depend on $Q$ will not contribute to these coefficients. 
	This includes the rate constants, which do not appear explicitly. 	
	Crucially, the coefficients $F_{\pm\rho}$s satisfy structural constraints related to the topology of the network.
	A graphical illustration of these structural relations is provided in Fig.~\ref{fig:Kirchhoff} for the particular case of a linear and a branched reaction pathway.  
	In the linear pathway, an arbitrary species $Z_i$ is transformed into a product species $Z_{i+1}$ by a reversible and unimolecular reaction $\rho_i$, and then $Z_{i+1}$ undergoes a similar reaction $\rho_{i+1}$. 
	In such case, both $j_{-\rho_i}$ and $j_{+\rho_{i+1}}$ depend solely on the concentration $z_{i+1}$ of species $Z_{i+1}$. 
	Consequently, $F_{-\rho_i}=F_{+\rho_{i+1}}$, because both terms are equal to $\dQ \log z_{i+1}$.
	For the branched pathway, a species $Z_0$ splits through the reaction $+ \rho_0$ into several products $Z_i$ with multiplicity $S_{-\rho_0}^i $, and Eq.~\eqref{eq:definition_F} leads to
	$F_{-\rho_0} = \sum_i S_{-\rho_0}^i \, F_{+\rho_i}$.
	
	For a general network, these structural relations take a form analogous to Kirchoff's laws (see Supplementary Note 3):
	\begin{equation}
		\label{eq:topology_constraints}
		\mathbf{F_-} = \mathbf{F_+} \cdot \left( \S_+^{-1} \cdot \S_- \right),
	\end{equation}
	provided $\S_+$ is invertible, which is the case in most networks of interest. 
	Note that the definition of the coefficients $F_{\pm \rho}$ and the structural constraints Eq.~\eqref{eq:topology_constraints} are valid even when $Q \neq Q^\ast$. 
	The structural constraints acting on the $F_{\pm\rho}$s play a key role in our framework, because these coefficients are intimately related to the affinities at the optimal current. 
	Indeed, when  $Q = Q^*$ one has:
	\begin{equation}
		\label{eq:relJF}
		j_{+\rho} \, F_{+\rho} = j_{-\rho} \,  F_{-\rho},
	\end{equation}
	because $\d_Q \, j^\ss_\rho = 0$.
	Using the definition of $\affrho$ and Eq.~\eqref{eq:relJF}, we obtain:
	\begin{equation}
		\label{eq:affinity_and_F}
		e^{\affrho} =  \dfrac{F_{-\rho}}{F_{+\rho}}.
	\end{equation}
	
	At this point of optimal current, Eq.~\eqref{eq:topology_constraints} defines a linear system of the form ${\M \cdot \mathbf{F_+} = 0}$,
	because the elements of  $\mathbf{F_-}$ can be expressed in terms of those of $\mathbf{F_+}$ by using the local affinities. 
	The solutions of this system will be trivial if $\M$, which contains information on both the topology and the affinities, is not singular. 
	Enforcing that the determinant of this matrix is zero results in a constraint involving solely the values of the optimal affinities and the elements of the stoichiometric matrices (see Supplementary Note 5).  
	It does not require an explicit evaluation of the steady-state rates or concentrations, nor does it involve kinetic parameters. 
	We illustrate this approach in the next section.
	
	The condition on the determinant is useful, but cannot easily be applied to large CRNs. 
	However, using Eq.~\eqref{eq:affinity_and_F}, the affinity of the overall reaction can be expressed only in terms of the $F_{\pm\rho}$ at the optimal flux:  
	\begin{equation}
		\label{eq:A_in_terms_of_F}
		\mathcal{A}^\ast  = \sum_{\rho} g_X^{\rho} \,  \log \left(  \dfrac{F_{-\rho}}{F_{+\rho}} \right).
	\end{equation} 
	Finding a  bound (a minimum or a maximum) for this global affinity corresponds to solving an optimization problem with Eq.~\eqref{eq:topology_constraints} acting as linear constraints and additional constraints due to tight coupling
	and the second law as detailed in the Methods section below (see Eq. \eqref{additional_constraints}).
	More details are provided in the Supplementary Note 11, where we establish conditions under which this optimization problem becomes concave.
	The solution to this optimization problem provides the thermodynamic bounds as well as information on the response coefficients using only the topology of the reaction network. As was the case with the method based on determinants, this link does not rely on an explicit evaluation of the steady-state fluxes or concentrations, nor does it require knowledge of kinetic parameters. In addition to this, the optimization approach can easily be used with large CRNs.

	
	\begin{figure*}
		\centering
		\includegraphics{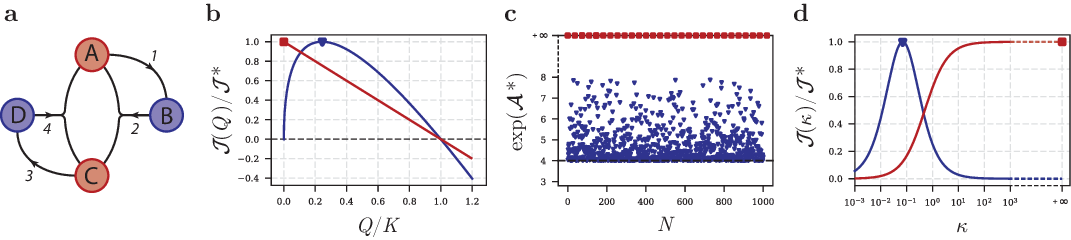}
		\caption{\label{fig:Hinshelwood_current}
			Optimal properties of the Hinshelwood cycle. \textbf{a} 
			Schematic representation of the Hinshelwood cycle with two types of species: red species (A or C) and blue species (B or D).
			\textbf{b} 
			Global flux normalized by its maximum value $\mathcal{J}^\ast$ as a function of $Q / K$ when a specific species of the cycle is chemostatted. 
			This flux has a zero-derivative maximum (blue triangle) if a blue species is chemostatted or simply reaches its maximal value at  $Q = 0$  (red square) if a red species is chemostatted.
			\textbf{c} 
			Cloud of points for the exponential of the global affinity $\mathcal{A}^*$, for randomly chosen sets of kinetic rate constants indexed by \textit{N}.   
			If a blue species is chemostatted, points are lower-bounded by $4$ (blue triangles) otherwise, the affinity diverges (red squares). 
			\textbf{d}
			Global flux normalized by its maximum value $\mathcal{J}^\ast$ as a function of the degradation rate $\kappa$ of a specific species. 
			When a blue species is degraded, a zero-derivative maximum exists and is reached at a finite value of $\kappa$, while if a red species is degraded, the global flux is monotonously increasing and reaches its maximal value at infinity. 
			Simulation parameters for \textbf{b} and \textbf{d} are $k_{+1} = k_{+3} = 1$ and $k_{+2} = k_{+4} =  k_{-1} = k_{-2} = k_{-3} = k_{-4} = 0.1$.  
		}
	\end{figure*}
	Since the bounds on the optimal affinity do not depend explicitly on the expressions of the steady-state concentrations or reaction rates, our method is applicable to large and complex networks, in which these concentrations and rates are too complex to be computed.
	For the same reason, the bounds hold even if the system features multistability, which is often found in autocatalytic networks \cite{Schuster2019}. 
	We  show this explicitly for the bistable Schl\"{o}gl  model \cite{Vellela2009} in the Supplementary Note 15.

	As an illustration of our approach, we derive lower bounds satisfied by the global affinity of the Hinshelwood autocatalytic cycle \cite{Hinshelwood1952} with intermediate species, as represented in  Fig.~\ref{fig:Hinshelwood_current}a. 
	When the intermediate species, \textsf{B} or \textsf{D}, are controlled, the constraints on the $F_{\pm \rho}$s can be satisfied and the global flux has a zero-derivative maximum (see the blue curve in Fig.~\ref{fig:Hinshelwood_current}b). 
	In that case, the affinity at the optimum of the current is lower-bounded, 
	\begin{equation}
		\mathcal{A}^\ast   \geq \ln 4,
	\end{equation}
	and the bound can be approached as closely as desired with an appropriate choice of rate constants, as shown in Figure \ref{fig:Hinshelwood_current}c (blue triangles).  
	When, instead, the concentration of \textsf{A} or \textsf{C} is maintained constant, the constraints acting on the $F_{\pm \rho}$s cannot be satisfied (see Supplementary Note 14).
	As a consequence, the global flux, $\Js$, has no zero-derivative maximum and, then, is a decreasing function of $Q$ whose largest value occurs at $Q = 0$ (see red curve in Fig.~\ref{fig:Hinshelwood_current}b), where the definition of the $F_{\pm \rho}$s and Eq.~\eqref{eq:relJF} do not apply. 
	As a result, $\exp (\mathcal{A}^\ast)$ diverges. 
	In addition to the Hinshelwood cycle, we also analyzed in the Supplementary Note 13 all the autocatalytic cores \cite{Blokhuis2020} ; the results are summarized in the Table \ref{tab:Recap_core}. 
	Importantly, the bound found for the five autocatalytic cores remain valid if unimolecular segments of reactions are added to any of these networks (see Supplementary Note 10 and Supplementary Figure 1). 
	
	\subsection*{Bounds for Type I and Type II networks} 
	
	
	Now, two situations arise depending on whether species $X$ appears or not as a reactant in the overall reaction. 
	When species  $X$  is a reactant in the overall equation, the reaction vector $\gX$ defines a seed-dependent mode of production \cite{Peng2022, andersen2021}. 
	In that case, the overall reaction is
	\begin{equation}
		\label{eq:Autocat_mode}
		\ce{$\alpha$\, \textsf{X} + $\mathrm{(\diamond)}$ -> $\beta$  \textsf{X} + $\mathrm{(\diamond)}$ }, 
	\end{equation}	
	with $\alpha$ and $\beta$ being integers such that $\beta > \alpha >0$, and ($\diamond$) represents all the spectator species. 
	Note that in our formalism, the concentrations of the external species are absorbed in the rate constants, so these species do not appear explicitly in the overall reaction. 
	The simplest network obeying Eq.~\eqref{eq:Autocat_mode} is a generalized version of the Type~I core presented in the example \eqref{eq:type_I_network} with an arbitrary number of intermediates species.
	As shown in the Supplementary Note 8, the global production rate in a generalized Type~I
			attains its maximum when 
			\begin{equation}
				\label{eq:Affinity_Generalized_Type_I}
				\mathcal{A}^\ast = \log \left(  \dfrac{\beta}{\alpha} \right).
			\end{equation}
			
			Let us now consider a specific class of networks that we call non-intersecting, for which the stoichiometric matrix of the products, $\S_-$,  is lower triangular up to permutations of species and reactions (see Supplementary Note 6). 
			This means that in non-intersecting networks, each reaction  only produces downstream species. We have shown that for non-intersecting networks, Eq.~\eqref{eq:Affinity_Generalized_Type_I} represents the lowest achievable bound for $\mathcal{A}^\ast$ when the overall reaction is seed-dependent. 
			We conjecture that this proposition is not restricted to non-intersecting networks but may hold in general. 

			Conversely, when species $X$ is not present as a reactant in the overall reaction, the reaction vector $\gX$ defines a seed-independent mode of production. 
			In that case, the overall reaction reads 
			\begin{equation}
				\label{eq:Allocat_mode}
				\ce{ $\mathrm{(\diamond)}$ -> $\beta~$  \textsf{X} + $\mathrm{(\diamond)}$ },
			\end{equation}
			where $\beta$ is a strictly positive integer. 
			It should be noted that Eq.~\eqref{eq:Allocat_mode} can describe overall reactions that are not necessarily derived from an autocatalytic network. 
			However, the presence of an optimal flux and the bounds on the corresponding affinity can only be guaranteed if there is an underlying autocatalytic network present.
			The simplest topology compatible with Eq.~\eqref{eq:Allocat_mode} is a generalization of the Type~II core: 
			\begin{equation}
				\label{eq:generalized_type_II}
				\ce{ 
					\alpha \, \textsf{X} 
					<-->
					$\textsf{Y}_{1}$
					<-->
					$\cdots$ 
					<-->
					$\textsf{Y}_{n}$ 
					<-->
					$\textsf{Y}_1$ + \beta \, \textsf{X} 
				}.
			\end{equation}
			This network is also non-intersecting (see Supplementary Note 9) and the global affinity associated to the maximum of $\mathcal{J}$ is
			\begin{equation}
				\label{eq:Affinity_Generalized_Type_II}
				\mathcal{A}^\ast = \log \left( \dfrac{\beta}{\alpha} + 1 \right).
			\end{equation}

			%
			%

			\section*{Discussion}
			
			We presented a theoretical framework that allows to derive constraints on the affinity and on the concentrations of autocatalysts, in the regime where all the fluxes in the network are tightly coupled and the rate of autocatalytic production is maximal. 
			In this specific regime and assuming mass-action law, the constraints can be derived using only the knowledge of the topology of the underlying reaction network but do not require a knowledge of the kinetics. 
			In this final section, we discuss potential applications of our work, its relation to non-equilibrium thermodynamics and potential extensions. 
			
			Besides the tight coupling condition, the core ingredient of our work is the presence of a maximum for the macroscopic flux with respect to a certain control parameter. 
			For simplicity, we derived our results assuming this parameter was directly related to a fixed concentration of one autocatalytic species, $Q$. 
			However, there exists flexibility in selecting this parameter, as long as the tight coupling property Eq.~\eqref{eq:steady_current} is preserved.
			In particular, tuning the concentration of a single autocatalyst, may not be very practical experimentally.
			A more manageable control parameter could be for instance a selective degradation using specific enzymes.
			For that case, we have checked that the property of tight coupling of the reaction fluxes is indeed preserved (see Eq.~\eqref{SM-eq:Strong_coupling_degradation} in the Methods section). 
			The macroscopic flux, $\Js$, becomes now function of a degradation rate constant $\kappa$ and becomes zero for $\kappa = 0$ at equilibrium, and above 
			a critical value $\kappa_\text{c}$ (provided $\kappa_\text{c} < \infty$), where degradation overcomes the production of the $X$ species by the autocatalytic network (see Eq.~\eqref{SM-eq:J_degradation} in the Methods section).
			This threshold $\kappa_\text{c}$ is experimentally accessible and has been considered before as a possible measure of fitness for autocatalytic networks \cite{Kolchinsky2021}.
			Importantly, because the constraints on the $F_{\pm \rho}$ and the steady-state are unaffected by the control procedure (chemostat or degradation), the bounds remain unchanged if one considers a degradation instead of chemostat.  
			We illustrated this point with the Hinshelwood cycle in Figure~\ref{fig:Hinshelwood_current}d:
			when \textsf{B} or \textsf{C} are being degraded, the current has a zero-derivative maximum verifying $\mathcal{A}^\ast \geq \log 4$. 
			While, on the contrary, when  \textsf{A} or  \textsf{D} undergo degradation, the current has no zero-derivative maximum and the overall affinity diverges. 
			
			
			The various bounds that we have encountered in this work measure how far from equilibrium a given network should be in order to deliver a maximal flux of autocatalytic production.  
			It is advantageous for a given network to operate close to this point not only because the global reaction is fast, but also because the system is then robust: at this point, variations in the concentration of the autocatalytic species are buffered and hardly affect the flux, which is thus stable.

			The affinity can also be interpreted as the entropy production associated with a steady production of species $X$ \cite{Polettini2014} or, equivalently, as the production of entropy per autocatalytic cycle \cite{Kolchinsky2021}.
			The bounds for the optimal affinities can thus be seen as the minimal cost required to maintain an optimal rate of autocatalytic production. 
			We also observed that this minimal cost increases dramatically when certain reactions within the network are at equilibrium or when certain key species are depleted, as we have illustrated in the case of the Hinshelwood cycle. 
			This is consistent with the idea that the thermodynamic cost should increase in a pathway when certain steps are at equilibrium \cite{Barenholz2017, Noor2014}. 
			
			Another natural extension of the present work would be to further investigate the thermodynamics of autocatalytic networks. It would be interesting  to explore possible connections between this work and a number of studies on thermodynamic trade-offs between dissipation, speed, and accuracy, and other recent studies on the response of non-equilibrium Markovian systems to perturbations \cite{Owen2023,liang2023,Aslyamov2023}.	
			
			
			In this work, we were mainly interested in chemical networks in a well-mixed environment assuming mass-action kinetics. These assumptions are valid for elementary reactions in well-mixed and dilute solutions, but might prove unrealistic in crowded environments and/or where chemical species interact, such as in  biological cells. 
			Alternative kinetic laws, like Michaelis-Menten kinetics, could be used in our framework as long as there is an elementary level at which the law of mass-action holds, and the coarse-graining from that microscopic level only involves unimolecular reactions. Thus, the assumption of mass-action law is an important limitation of our work, but it is the price to pay to obtain bounds which are independent of the kinetics. 
			Similarly, a recent work reported a method to compute concentration ranges for metabolites in the absence of knowledge from kinetics but the method is only applicable for mass-action law networks in the limit of dilute systems \cite{kuken2019}.

			Our framework could be used as a computational approach 
			to predict concentration ranges and thermodynamic affinities from large scale metabolic models,	
			which could be useful for the design of autocatalytic or self-replicating systems \cite{Barenholz2017}.
			It could also be used to provide information on the topological complexity of observed chemical networks. Indeed, we observed that the bounds tend to increase with the topological complexity of the network, as shown in Table~\ref{tab:Recap_core}. One measure of topological complexity is the connectivity of the network, and precisely, we have shown that the bound increases with the degree of branching of a reaction when the network contains only one such reaction (see Supplementary Note 7). 
			A generalization beyond this case may be within reach.

			Finally, we note that autocatalytic systems are of particular importance for scenarios on the origin of life. Autocatalytic networks can amplify initially small numbers of molecules and the rate at which such species are produced would certainly play a role in the competition between molecules. In this context, running at an optimal and robust production rate could provide an autocatalytic network with a significant advantage in terms of chemical selection. It would thus be interesting to further explore consequences of this framework for assessing the robustness and the evolvability of autocatalytic networks. 
			
			\section*{Methods}
			\NiceMatrixOptions{
				code-for-last-row = \color{black!70}\scriptstyle,
				code-for-first-col = \color{black!70}\scriptstyle,
			}
			\subsection*{Setup and notation}
			We assume that the chemical reaction network (CRN) is described by a stoichiometric matrix of the form 
			\begin{equation}
				\label{SM-eq:Nabla}
				\boldsymbol{\nabla} = \quad \begin{pNiceArray}{w{c}{1em}c:w{c}{1em}c}[last-row, first-col]
					& \Block{2-4}{\Sex} &  &  & \\
					& & & & \\
					\hdottedline
					\Vdots[line-style={solid, <->}, shorten=2pt]_{\small \rotatebox{90}{$\mathcal{E}$} } & \Block{2-2}{\S} &  & \Block{2-2}{ \mathbf{0} } &   \\
					& & & & \\
					& \Ldots[line-style={solid, <->}, shorten=5pt]_{ \mathcal{P} } \\
				\end{pNiceArray},
				\vspace*{0.5em}
			\end{equation}
			where $\E$ is the set of all species and $\P$ is the set of all reactions in the full system. 
			The restriction of $\boldsymbol{\nabla}$ on $\Eex$ (denoted $\Sex$) describes external species, while $\S$ is the stoichiometric matrix of autocatalytic species. By assumption, it is a square invertible matrix.
			Thus, one has $\E = \Eex \cup \Z$ and $\P = \Pex \cup \R$, where $\Z$ is the set of autocatalytic species and $\R$ is the set of autocatalytic reactions. 
			The autocatalytic network is coupled to external species
			in $\Eex$ which are involved in additional reactions $\Pex$.  
			Note that a somewhat similar matrix decomposition has already been introduced to study geometrical features of non-equilibrium reaction networks \cite{Cengio2022}.

			After chemostatting all the external species, the autocatalytic CRN is still able to reach detailed balance. 
			To see this, one can multiply the matrix $\boldsymbol{\nabla}$ on its left hand side by the row vector $\left( \boldsymbol{\mu}_\text{ex},~ \boldsymbol{\mu} \right)$ where $\boldsymbol{\mu}_\text{ex}$ are the chemical potentials of the external species and $\boldsymbol{\mu}$ the chemical potentials of the autocatalytic species. 
			This yields $\boldsymbol{\mu}_\text{ex} \cdot \left( \Sex \right)_\R + \boldsymbol{\mu} \cdot \S $, with $\left( \Sex \right)_\R$ being the restriction of $\Sex$ to the space of autocatalytic reactions $\R$.
			As $\S$ is non-singular, a solution that satisfies detailed balance always exists:
			\begin{equation}
				\label{SM-eq:Balance_chemical_potential}
				\boldsymbol{\mu}^\mathrm{eq} = - \boldsymbol{\mu}_\text{ex} \cdot \left( \Sex \right) _\R \cdot \S^{-1}. 
			\end{equation}
			This means that the autocatalyic CRN is able to reach an equilibrium state even though the additional reactions $\Pex$ are kept away from equilibrium ($\boldsymbol{\mu}_\text{ex} \cdot (\Sex)_{\Pex } \neq \mathbf{0}$).

			\subsection*{Additional constraints}
			
			Applying the second law of thermodynamics on Eq.~\eqref{eq:EPR_main} imposes that the global flux is positive for $Q \in [0, K[$ and negative for $Q \geq K$ thus, $Q^\ast$ lies in $[0, K[$, where $ \Js (Q^\ast) > 0 $.
			Then, because of the tight coupling condition Eq.~\eqref{eq:steady_current}, the signs of $j_\rho $ and $g_X^\rho$ are the same.   
			Further, if $g_X^\rho > 0 $ then $j_\rho > 0$ and thus $\mathcal{A}_\rho > 0$; conversely, if $g_X^\rho < 0 $ then $j_\rho < 0 $, yielding $\mathcal{A}_\rho < 0 $.
			We are then left with the special and important case where $g_X^\rho = 0$, which implies that both $j^\ss_\rho$ and  $\mathcal{A}_\rho = 0$ vanish. 
			In that case, the corresponding reaction $\rho$ is at equilibrium.
			To summarize :
			\begin{equation}
				\label{additional_constraints}
				\forall \rho \in \mathcal{R}, ~ g_X^\rho \, \mathcal{A}_\rho > 0, \quad 
				{\rm and} \quad \mathcal{A}_\rho= 0 
				\quad {\rm when} \quad g_X^\rho = 0.
			\end{equation}
			Note that these conditions directly translate into additional non-linear constraints on the $F_{\pm \rho}$s at the optimum.
			
			\subsection*{Rescaled modes of production}
			
			Because $\S$ is integer-valued, there exists a smallest ${n \in \mathbb{N}^\ast}$ such that $n \, \S^{-1}$ is also integer-valued.
			The columns of  $n \, \S^{-1}$ define the rescaled modes of production.
			From Eq.~\eqref{eq:overall_affinity_main}, along these rescaled modes, the global affinity is $ n \, \mathcal{A}$.
			Since the value of each of the elementary fluxes needs to remain unchanged, the tight coupling condition Eq.~\eqref{eq:steady_current} implies that the macroscopic current should be $\Js / n$ after the rescaling.
			As a result, from Eq.~\eqref{eq:EPR_main}, the EPR is preserved by the rescaling. \\ 
			In Eqs.~\eqref{eq:Autocat_mode} and \eqref{eq:Allocat_mode}, we implicitly used the rescaled modes of production so that $\alpha$, $\beta \in \mathbb{N}$.

			\subsection*{Extension to the case of specific degradation}
			
			We show here that the bounds derived by considering that an autocatalytic species is chemostatted remain valid when the chemostatting procedure is replaced by a specific degradation of the same species. 
			Let us call $X$ the autocatalytic species in question.  
			We can introduce an augmented stoichiometric matrix and an augmented flux vector to take into account the degradation: 
			\begin{align}
				\S^\prime & =  \begin{pNiceArray}{w{c}{0.5cm}:c}
					\mathbf{S}^X & -1 \\[1em] 
					\S^Y & \mathbf{0} 
				\end{pNiceArray}, & \quad & 
				\mathbf{v} = \begin{pNiceMatrix} \mathbf{j} \\[1em] \kappa \, f(x) \end{pNiceMatrix},
			\end{align}
			where $\kappa$ is non-negative. 
			The degradation rate is described by the function $f(x)$, which can be a simple power law $x^n$ with $n>0$, or a more sophisticated expression, such as a Hill function:
			\begin{equation}
				f(x)=\frac{x^n}{x^n + K},
			\end{equation}
			in which $K$ is usually referred to as the apparent dissociation constant. 
			The Hill function is often used to model kinetics involving the fixation of a substrate on macromolecules (such as proteins), and includes the Michaelis-Menten law as a special case (\emph{i.e.} $n=1$). 
			The dynamics of such an extended system obeys  
			\begin{equation}
				\label{SM-eq:degradation_dynamics}
				\d_t \mathbf{z} = \S^\prime \cdot \mathbf{v}.
			\end{equation}
			A steady-state of this new system consists of $\mathbf{v} \in \ker \left[ \S^\prime \right] $, the latter being spanned by $\gX^\prime = \left( \gX, ~ 1 \right)^\top$, 
			\begin{equation}
				\label{SM-eq:SS_eqn_degradation}
				\S^\prime \cdot \gX^\prime = \begin{pmatrix}  \mathbf{S}^X \cdot \gX - 1 \\[5pt]   \S^Y \cdot \gX \end{pmatrix} = \mathbf{0}.
			\end{equation}
			Consequently, the steady elementary fluxes associated with  Eq.~\eqref{SM-eq:degradation_dynamics} are proportional to $\gX^\prime$, implying the tight-coupling condition:
			\begin{equation}
				\label{SM-eq:Strong_coupling_degradation}
				\mathbf{j}^\ss = \Js (\kappa) ~ \gX.
			\end{equation}
			The steady-state fluxes of the various reactions still follow the law of mass action Eq.~\eqref{eq:mass_action_kin}, but are now parameterized by $\kappa$ instead of $Q$. 
			Consequently, we recover the definition of the $F_{\pm \rho}$s as the log-derivatives of the elementary fluxes Eq.~\eqref{eq:definition_F}.
			From that, the structural constraints Eq.~\eqref{eq:topology_constraints} follow. 
			
			At steady-state, the tight coupling condition Eq.~\eqref{SM-eq:Strong_coupling_degradation} implies:
			\begin{equation}
				\label{SM-eq:J_degradation}
				\Js (\kappa) = \kappa \, f(x). 
			\end{equation}
			Hence, the global flux necessarily vanishes when $\kappa = 0$ because the system reaches equilibrium in the absence of degradation. 
			Additionally, Eq.~\eqref{SM-eq:J_degradation}  implies that ${\Js (\kappa) \geq 0}$, which allows us to recover the additional constraints derived above.  
			We now introduce $\kappa_\text{c} \in \, ]0, + \infty]$, which is defined as the value of the degradation rate such that $x (\kappa_\text{c}) = 0$.  
			If $\kappa_\text{c} < +\infty$ then, for any $\kappa \geq \kappa_\text{c}$, $x( \kappa ) = 0 $ is the only physically acceptable solution, so that Eq.~\eqref{SM-eq:J_degradation} implies $\Js ( \kappa ) = 0$ as well. 
			On the other hand, if $\kappa_\text{c}$ diverges, Eq.~\eqref{SM-eq:J_degradation} imposes that
			\begin{equation}
				\label{SM-eq:X_species_degradation}
				x (\kappa)^n \underset{+ \infty}{\sim} \kappa^{- \alpha}, ~ \alpha \geq 1. 
			\end{equation}
			From that, two qualitatively different situations can be found. 
			If $\alpha > 1$, the global flux tends to vanish as $\kappa$ goes to infinity: $ \Js ( + \infty )  = 0$. 
			Instead, if $\alpha = 1$, the global flux converges to a finite non-zero value, $ \Js ( + \infty )  > 0$. 
			Such systems are similar to those having a non-vanishing global flux when $Q = 0$ in the chemostatted case. 
			As before, if the constraints are incompatible,  $\Js ( \kappa )$ is a monotonous function, attaining its maximum when the $X$ species is completely depleted as shown by the red curve in Fig.~\ref{fig:Hinshelwood_current}d.
			In the system with specific degradation, this corresponds to impose a diverging degradation rate, $\kappa^\ast = \kappa_\text{c} = + \infty$.
			
			\section*{Data Availability}
			
			No datasets were generated or analyzed in this study.

			\section*{Code Availability}
			
			The code that generated the plots is available from the corresponding author upon reasonable request.

			\section*{Acknowledgments}
			
			We acknowledge J. Unterberger, L. Jullien, W. Liebermeister, P. Gaspard  and A. Blokhuis for stimulating discussions. D.L. received support from the grants ANR-11-LABX-0038, ANR-10-IDEX-0001-02.
			
			\section*{Authors Contributions}

			Y.D.D., D.L. designed the research and received funding. 
			A.D. performed the research. 
			A.D., Y.D.D., D.L. wrote the paper. 
			
			\section*{Competing Interest}
			
			The authors declare no competing interests.

			\section*{Additional Information}

			\paragraph{Supplementary information}

			\paragraph{Correspondence} and requests for materials should be addressed to Armand Despons.

			\NiceMatrixOptions{
				code-for-last-row = \color{black!70}\scriptstyle,
				code-for-first-col = \color{black!70}\scriptstyle,
			}
			
			\newcommand{\SI}{\begin{pNiceMatrix}[last-row, first-col]
					\textsf{A} & -1  & 2  \\
					\textsf{B} & 1   & -1 \\
					& \mathsfit{1}   & \mathsfit{2}  \\
				\end{pNiceMatrix}
			}
			\newcommand{\GI}{\begin{pNiceMatrix}[last-row, first-col] 
					\mathsfit{1} & 1 & 2 \\ 
					\mathsfit{2} & 1 & 1 \\
					& \mathbf{g}_\textsf{A}   & \mathbf{g}_\textsf{B}  \\
				\end{pNiceMatrix}
			}
			
			\newcommand{\SII}{\begin{pNiceMatrix}[last-row, first-col]  
					\textsf{A} & -1 & 0 & 1 \\ 
					\textsf{B} &  1 & -1 & 1 \\ 
					\textsf{C} & 0 & 1 & -1 \\
					& \mathsfit{1}  & \mathsfit{2}  & \mathsfit{3} \\
				\end{pNiceMatrix}
			}
			\newcommand{\GII}{\begin{pNiceMatrix}[last-row, first-col]  
					\mathsfit{1} & 0 & 1 & 1 \\ 
					\mathsfit{2} & 1 & 1 & 2 \\ 
					\mathsfit{3} & 1 & 1 & 1 \\ 
					& \mathbf{g}_\textsf{A}   & \mathbf{g}_\textsf{B} & \mathbf{g}_\textsf{C} \\
				\end{pNiceMatrix}
			}
			
			\newcommand{\SIII}{\begin{pNiceMatrix}[last-row, first-col]   
					\textsf{A} & -1 & 0 & 1 \\
					\textsf{B} & 0 & -1 & 1 \\
					\textsf{C} & 1 & 1 & -1 \\ 
					& \mathsfit{1}  & \mathsfit{2}  & \mathsfit{3} \\
			\end{pNiceMatrix}}
			\newcommand{\GIII}{\begin{pNiceMatrix}[last-row, first-col]  
					\mathsfit{1} & 1 & 1 & 1 \\ 
					\mathsfit{2} & 1 & 0 & 1 \\ 
					\mathsfit{3} & 1 & 1 & 0 \\ 
					& \mathbf{g}_\textsf{A}   & \mathbf{g}_\textsf{B} & \mathbf{g}_\textsf{C} \\
				\end{pNiceMatrix}
			}
			
			\newcommand{\SIV}{\begin{pNiceMatrix}[last-row, first-col]   
					\textsf{A} & -1 & 1 & 0 \\
					\textsf{B} & 1 & -1 & 1 \\
					\textsf{C} & 1 & 1 & -1 \\ 
					& \mathsfit{1}  & \mathsfit{2}  & \mathsfit{3} \\
			\end{pNiceMatrix}}
			\newcommand{\GIV}{\begin{pNiceMatrix}[last-row, first-col]  
					\mathsfit{1} & 0 & 1/2 & 1/2 \\ 
					\mathsfit{2} & 1 & 1/2 & 1/2 \\ 
					\mathsfit{3} & 1 & 1 & 0 \\ 
					& \mathbf{g}_\textsf{A}   & \mathbf{g}_\textsf{B} & \mathbf{g}_\textsf{C} \\
				\end{pNiceMatrix}
			}
			
			\newcommand{\SV}{\begin{pNiceMatrix}[last-row, first-col]   
					\textsf{A} & -1 & 1 & 1 \\
					\textsf{B} & 1 & -1 & 1 \\
					\textsf{C} & 1 & 1 & -1 \\ 
					& \mathsfit{1}  & \mathsfit{2}  & \mathsfit{3} \\
			\end{pNiceMatrix}}
			\newcommand{\GV}{\begin{pNiceMatrix}[last-row, first-col]  
					\mathsfit{1} & 0 & 1/2 & 1/2 \\ 
					\mathsfit{2} & 1/2 & 0 & 1/2 \\ 
					\mathsfit{3} & 1/2 & 1/2 & 0 \\ 
					& \mathbf{g}_\textsf{A}   & \mathbf{g}_\textsf{B} & \mathbf{g}_\textsf{C} \\
				\end{pNiceMatrix}
			}
			
			\newpage
			
			\begin{table*}[t]
				\centering
				\SetTblrInner{vspan=even}
				\begin{tblr}{width=0.5\textwidth, colspec={Q[c, c, wd=3cm]|c|c|Q[c, c, wd=5cm]|Q[c, c, wd=2cm]},
					}
					{\Large Motif} & {\Large $\S$} & {\Large $\S^{-1}$} & {\Large Overall equation along $\gX$} & {\Large Bound on $\mathcal{A}^\ast$} \\
					\hline
					\hline
					\SetCell[r=2]{c} {\includegraphics[trim={0.1cm 0 0.1cm 0}, clip]{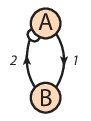} \\ \textsf{{\large Type~I}}} & \SetCell[r=2]{c} $\SI$ & \SetCell[r=2]{} $\GI$ & \ce{\textsf{A} + \textsf{B} ->[$\mathbf{g}_\textsf{A}$] 2\textsf{A} + \textsf{B}} & $\mathcal{A}^\ast = \log 2$ \\ 
					&                        &                       &  \ce{2\textsf{A} + \textsf{B} ->[$\mathbf{g}_\textsf{B}$] 2\textsf{A} + 2\textsf{B}} & $\mathcal{A}^\ast \geq \log 2$ \\
					\hline
					\SetCell[r=3]{c} {\includegraphics[trim={0.1cm 0 0.1cm 0}, clip]{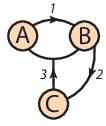} \\ \textsf{{\large Type~II}}}    & \SetCell[r=3]{c} $\SII$ & \SetCell[r=3]{} $\GII$ & \ce{\textsf{B} + \textsf{C} ->[$\mathbf{g}_\textsf{A}$] \textsf{A} + \textsf{B} + \textsf{C}} & $\mathcal{A}^\ast = \log 2$  \\
					&                         &                        &  \ce{\textsf{A} + \textsf{B} + \textsf{C} ->[$\mathbf{g}_\textsf{B}$] \textsf{A} + 2\textsf{B} + \textsf{C}} & $\mathcal{A}^\ast \geq \log 2$  \\
					C													   &                         &                        &  \ce{\textsf{A} + 2\textsf{B} + \textsf{C} ->[$\mathbf{g}_\textsf{C}$] \textsf{A} + 2\textsf{B} + 2\textsf{C}} & $\mathcal{A}^\ast \geq \log 2$  \\
					\hline
					\SetCell[r=3]{c} {\includegraphics[trim={0cm 0 0cm 0}, clip]{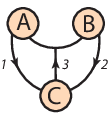} \\ \textsf{{\large Type~III}}} & \SetCell[r=3]{c} $\SIII$ & \SetCell[r=3]{} $\GIII$ & \ce{\textsf{A} + \textsf{B} + \textsf{C} ->[$\mathbf{g}_\textsf{A}$] 2\textsf{A} + \textsf{B} + \textsf{C}} & $\mathcal{A}^\ast \geq \log 2$  \\ 
					&                          &                         & \ce{\textsf{A} + \textsf{C} ->[$\mathbf{g}_\textsf{B}$] \textsf{A} + \textsf{B} + \textsf{C}} & $\mathcal{A}^\ast \geq \log 2$  \\ 
					&                          &                         & \ce{\textsf{A} + \textsf{B} ->[$\mathbf{g}_\textsf{C}$] \textsf{A} + \textsf{B} + \textsf{C} } & $\mathcal{A}^\ast \geq \log 2$  \\
					\hline 
					\SetCell[r=3]{c} {\includegraphics[trim={0cm 0.8cm 0.1cm 0}, clip]{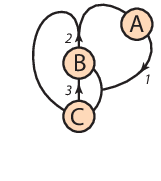} \\ \textsf{{\large Type~IV}}} & \SetCell[r=3]{m} $\SIV$ & \SetCell[r=3]{m} $\GIV$ & \ce{\textsf{B} + \textsf{C} ->[$\mathbf{g}_\textsf{A}$] \textsf{A} + \textsf{B} + \textsf{C}} &  $\mathcal{A}^\ast \geq \log 3$   \\ 
					&                         &                         & \ce{\frac{1}{2} \textsf{A} + \frac{1}{2} \textsf{B} + \textsf{C} ->[$\mathbf{g}_\textsf{B}$] \frac{1}{2} \textsf{A} + \frac{3}{2} \textsf{B} + \textsf{C}} & $\mathcal{A}^\ast \geq \dfrac{\log 3}{2}$  \\ 
					&                         &                         & \ce{\frac{1}{2} \textsf{A} + \frac{1}{2} \textsf{B} ->[$\mathbf{g}_\textsf{C}$] \frac{1}{2} \textsf{A} + \frac{1}{2} \textsf{B} + \textsf{C}} & $\mathcal{A}^\ast \geq \dfrac{\log 3}{2}$   \\
					\hline 
					\SetCell[r=3]{c} {\includegraphics[trim={0cm 0cm 0cm 0cm}, clip]{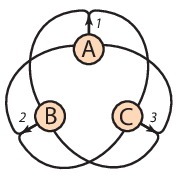} \\ \textsf{{\large Type~V}}} & \SetCell[r=3]{c} $\SV$   & \SetCell[r=3]{} $\GV$   & \ce{\frac{1}{2} \textsf{B} + \frac{1}{2} \textsf{C} ->[$\mathbf{g}_\textsf{A}$] \textsf{A} + \frac{1}{2} \textsf{B} + \frac{1}{2} \textsf{C}} &  $\mathcal{A}^\ast \geq \log 3$  \\
					&                         &                         &  \ce{\frac{1}{2} \textsf{A} + \frac{1}{2} \textsf{C} ->[$\mathbf{g}_\textsf{B}$] \frac{1}{2} \textsf{A} + \textsf{B} + \frac{1}{2} \textsf{C}} &  $\mathcal{A}^\ast \geq \log 3$   \\ 
					&                         &                         & \ce{\frac{1}{2} \textsf{A} + \frac{1}{2} \textsf{B} ->[$\mathbf{g}_\textsf{C}$] \frac{1}{2} \textsf{A} + \frac{1}{2} \textsf{B} + \textsf{C}} & $\mathcal{A}^\ast \geq \log 3$ \\
					\hline
					\hline
				\end{tblr}
				\vspace*{1\baselineskip}
				\caption{\label{tab:Recap_core} 
					Bounds on the chemical affinities at the maximum of the macroscopic flux for the five autocatalytic cores \cite{Blokhuis2020}. 
				}
			\end{table*}
			
		\newpage
		\onecolumngrid
		
		\renewcommand{\theequation}{S\arabic{equation}}

		%
		%
		\section*{
			Supplementary Note 1:
			General properties of the autocatalytic CRN}
		
		We assume that external species are externally controlled (chemostatted) and include their concentrations in the rate constants of the reactions.  
		For any $\rho \in \R$:
		\begin{equation}
			\ce{ $\sum_{\sigma} S^{\sigma}_{+\rho} \, Z_{\sigma} $ <-->[$k_{+\rho}$][$k_{-\rho}$] $ \sum_{\sigma}   S^{\sigma}_{-\rho} \, Z_{\sigma} $},
		\end{equation}
		with the autocatalytic  species $Z_\sigma \in \Z$ being indexed by $\sigma$. The $k_{\pm \rho}$s are the (effective) kinetic rate constants and the coefficients $S^\sigma_{+\rho}$ (resp. $S^\sigma_{-\rho}$) encode the molecularities
		of the product for the forward (resp. reverse) transition $+\rho$ (resp. $-\rho$), such that:
		\begin{equation}
			\label{SM-eq:stoich_matrix}
			\mathbb{S} = \mathbb{S}_- - \mathbb{S}_+. 
		\end{equation}
		The deterministic dynamics of the full CRN is given by
		\begin{align}
			\d_t \boldsymbol{e} & = \Sex \cdot \boldsymbol{v} + \boldsymbol{I}^{ex} = 0 \label{SM-eq:Clamp_ex_species} \\
			\d_t \boldsymbol{z} & = \S \cdot \boldsymbol{j} \label{SM-eq:Dyn_internal_CRN},
		\end{align}
		where $\boldsymbol{e}$ are the concentrations of the external species, $\boldsymbol{v}$ are the net fluxes of all the reactions ($\P$), $\boldsymbol{j}$ its restriction on the autocatalytic reaction ($\R$): $\boldsymbol{j} \equiv \boldsymbol{v}^{\R}$ and $\boldsymbol{I}^{ex}$ the external flux. \\
		We call respectively $\C$ and $\L$ the set of cycles and the set of conservation laws associated to $\boldsymbol{\nabla}$ (\emph{i.e.} its kernel and its cokernel). 
		Similarly, we refer to $\widetilde{\C}$ and $\widetilde{\L}$ respectively as the cycles and the conservation laws of $\boldsymbol{\nabla}^\Z$, the restriction of $\boldsymbol{\nabla}$ on the set of autocatalytic  species. 
		The rank-nullity theorem links the number of external species, $| \Eex |$, to the number of emergent cycles (cycles in $\widetilde{\C}$ but not in $\C$), $ | \widetilde{\C} |  - | \C |$, and the number of broken conservation laws, $| \L | - | \widetilde{\L} |$: 
		\begin{equation} 
			\label{SM-eq:Rank_nullity_thm}
			|\Eex| =  | \widetilde{\C} |  - | \C | +  | \L | - | \widetilde{\L} |. 
		\end{equation}
		Additionally, the non-singularity of $\S$ dictates that the only steady state for the autocatalytic  dynamics is the equilibrium configuration $\boldsymbol{j}_\text{eq} = \boldsymbol{0}$. 
		Nonetheless, depending on how the external species are coupled to the autocatalytic CRN, two situations can emerge, impacting \eqref{SM-eq:Rank_nullity_thm}.

		\paragraph{Case $\Pex = \varnothing$}
		
		If the external species are not involved in any additional reaction, the full stoichiometric matrix simplifies to 
		\begin{equation}
			\label{SM-eq:Full_stoichiometry_I}
			\boldsymbol{\nabla} = \quad \begin{pNiceArray}{w{c}{0.5cm}c}[last-row, first-col]
				& \Block{2-2}{\Sex} &  \\
				& &    \\
				\hdottedline
				\Vdots[line-style={solid,<->}]_{\small \rotatebox{90}{$\mathcal{E}$} } & \Block{2-2}{\S}  & \\
				&  &  \\
				&  \Ldots[line-style={solid,<->}, shorten=6pt]_{ \mathcal{R} } \\
			\end{pNiceArray},
			\vspace*{0.5em}
		\end{equation}
		where we see that the set of autocatalytic reaction $\mathcal{R}$ is shared by $\S$ and $\Sex$. 
		In that case, the matrix encoding the properties of the open CRN simplifies as well: $\boldsymbol{\nabla}^\Z = \S$. 
		Hence the non-singularity of $\S$ implies that: 
		\begin{enumerate}
			\item $\widetilde{\C} = \C = \varnothing$, no emergent cycle appears after chemostating the external species,  
			
			\item $\widetilde{\L} = \varnothing$, the open CRN has no conservation law.
		\end{enumerate}
		The absence of conservation in $\S$ is a hallmark of autocatalysis. However, chemical consistency enforces (at least) the conservation of the total mass when the full network is closed ($\boldsymbol{I}^{ex}=\boldsymbol{0}$) and as consequence, $\L$ is never empty. 
		It follows from \eqref{SM-eq:Rank_nullity_thm} that the number of conserved quantities in the full network is equal to the number of external species, $| \L | = | \Eex |$. 
		Thus, all the conservation laws are broken once the species in $\Eex$ are chemostatted, which means that no emergent cycle emerges.
		Using the classification of Anvizini \emph{et al.} \cite{Avanzini2021}, in that case the external species are \emph{potential species} that are not able to break detailed balance. 
		As a consequence, for any values of the concentrations of external species, nothing prevents the autocatalytic CRN from reaching equilibrium. 
		
		\paragraph{Case $\Pex \neq \varnothing$}
		
		There is no specific reason to assume that external species are not involved in additional reactions (for example ATP hydrolysis in the context of metabolic networks).
		After clamping the external species, the stoichiometric matrix of the open system is: 
		\begin{equation}
			\boldsymbol{\nabla}^\Z \equiv ~\quad \label{SM-eq:Nabla} \begin{pNiceArray}{cc:cc}[last-row, first-col]
				\Vdots[line-style={solid, <->}, shorten=0pt]_{\small \rotatebox{90}{$\mathcal{Z}$} } & \Block{2-2}{\S} &  & \Block{2-2}{ \boldsymbol{0} } &   \\
				& & & & \\
				& \Ldots[line-style={solid, <->}, shorten=5pt]_{ \mathcal{P} } \\
			\end{pNiceArray},
			\vspace*{0.5em}
		\end{equation}
		which has  $| \Pex |$ independent (emergent) cycles.
		Therefore, still following the nomenclature proposed by Anvizini, there will be $| \Pex |$ \emph{force species}, which can possibly break detailed balance. \\
		Because the autocatalytic  species are not involved in reactions $\Pex$, this subset of reactions can remain out of equilibrium while the autocatalytic CRN is still able reach detailed balance. 
		However, as seen in the first section of the Methods section in the main text, the autocatalytic CRN is able to reach an equilibrium state nonetheless.

		\section*{
			Supplementary Note 2:
			Modes of production of the network}
		
		As we just saw, chemostating the external species is not sufficient to maintain autocatalysis in a steady-state because the autocatalytic CRN will reach equilibrium. 
		Consequently, maintaining autocatalysis steadily requires that the concentration of (at least) one of the autocatalytic species be chemostatted as well. We will denote this chemostatted autocatalytic  species with an $X$, which thus has a  constant concentration $x$. 
		As in the main text, the $|\Z|-1$ non-chemostatted autocatalytic  species are called the $Y$-species, and their concentration are grouped in $\boldsymbol{y}$.

		Taking this into account in \eqref{SM-eq:Dyn_internal_CRN} leads to Eqs.~(15)-(16)  of the main text for the dynamics of the out-of-equilibrium autocatalytic CRN.
		The inverse of $\S$ splits accordingly: 
		\begin{equation}
			\S^{-1} = \begin{pNiceArray}{c c} \gX, & \mathbb{G}_Y  \end{pNiceArray},
		\end{equation}
		where we stacked in $\mathbb{G}_Y$ the columns of  $\S^{-1}$ associated to the $Y$ species. Consequently, 
		\begin{equation}
			\S \cdot \S^{-1} = \begin{pNiceArray}{c:w{c}{1.2cm}}
				\mathbf{S}^X \cdot \gX & \mathbf{S}^X \cdot \mathbb{G}_Y \\
				\Hdotsfor{2} \\[0.5em]
				\S^Y \cdot \gX & \S^Y \cdot \mathbb{G}_Y \\[0.5em]
			\end{pNiceArray} = \text{Id.},
		\end{equation}
		in which $\mathbf{S}^X$ and $\S^Y$ were respectively defined as the row vector of $\S$ associated to the chemostated autocatalytic species and the restriction on the $Y$-species of $\S$. 
		This leads to:
		\begin{align} 
			\mathbf{S}^X \cdot \gX & = 1 \label{SM-eq:Elementary_mode_1} \\
			\S^Y \cdot \gX & = \boldsymbol{0} \label{SM-eq:Elementary_mode_2}. 
		\end{align}	
		These two equations quantify the two properties of the \emph{elementary mode of production} of species $\textsf{X}$: $\gX$ is reaction pathway that produces one extra unit of species $\textsf{X}$ \eqref{SM-eq:Elementary_mode_1} letting the $Y$-species unchanged (\eqref{SM-eq:Elementary_mode_2}). 
		Additionally, from \eqref{SM-eq:Elementary_mode_2} and the rank-nullity theorem using $\widetilde{\C} = \varnothing$, we have $\ker \left( \S^Y \right) = \mathrm{span} \left( \gX \right)$. \\
		
		\paragraph{Seed-dependency of $\gX$}
		Let us note $\R_\textsf{X} \subset \R$ the subset of autocatalytic reactions having species $\textsf{X}$ as reactant, note that if $\rho$ has \textsf{X} as product then $- \, \rho \in \mathcal{R}_\textsf{X}$. 
		The sign of the restriction of $\gX$ to $\R_\textsf{X}$, $\gX^{\R_\textsf{X}}$, allows to know if $\gX$ is a seed-dependent or a seed-independent mode of production. 
		If $\gX^{\mathcal{R}_\textsf{X}}$ has at least one positive component, \textsf{X} is consumed by, at least, one reaction in the overall reaction consequently, the mode of production will be seed-dependent.
		Conversely, if $\gX^{\R_\textsf{X}} \leq 0 $ (component-wise), \textsf{X} will not be present in the reactant side in the overall equation then $\gX$ will be seed-independent. \\
		With that we can show that, along a seed-dependent mode of production, the macroscopic current is necessarily vanishing at zero. 
		Indeed, if $\gX$ is seed-dependent there exists $\rho \in \R_\textsf{X}$ such that, $ g_\textsf{X}^{\rho} > 0$. 
		Hence, for all $ \, Q \, \in \,  [0, K], ~ j_{\rho}^\ss \left( Q \right)  \geq 0$.
		In addition, $j^\ss_{\rho} \left( 0 \right)  = - j^\ss_{- \rho} \left( 0 \right)  \leq 0$. Combining this with the previous inequality, we get $ j^\ss_{\rho} \left( 0 \right)  = 0$.
		Because $\Js ( Q )  = (g_\textsf{X}^{\rho})^{-1} \, j^\ss_{\rho} \left( Q \right) $, we obtain that  $\Js  (0)  = 0$.

		%
		%
		\section*{
			Supplementary Note 3:
			Structural constraints}
		
		At the steady state, the concentrations of the $Y$-species relax to their $Q$-dependent values $\boldsymbol{y}^\ss (Q)$, 
		$Q$ stands for the concentration of species \textsf{X}. 
		Isolating the concentration of the chemostatted species in the steady currents, we can write: 
		\begin{equation}
			\label{SM-eq:MAL_ss}
			j_{\pm \rho}^\ss \left(Q\right) = k_{\pm \rho } \, Q^{S^X_{\pm \rho} } \prod_{Y_\sigma \in Y} {y}_\sigma(Q)^{\left({S^Y}\right)^\sigma_{\pm \rho}}.
		\end{equation}
		The total derivative of \eqref{SM-eq:MAL_ss} reads 
		\begin{align}
			\dQ j_{\pm \rho}^\ss & = \partial_Q j_{\pm \rho}^\ss + \d_Q \boldsymbol{y} \cdot \partial_{\boldsymbol{y}} j_{\pm \rho}^\ss 
			= \left( \dfrac{S^X_{\pm \rho}}{Q} + \sum_{Y_\sigma \in Y} \left({S^Y}\right)^\sigma_{\pm \rho} \, \dfrac{\d_Q y_\sigma}{y_\sigma} \right) \,  j_{\pm \rho}.
		\end{align} 
		Consequently, the $F_{\pm \rho}$s are defined as 
		\begin{align}
			\label{SM-eq:F_pm}
			\boldsymbol{F}_\pm  = \dQ \boldsymbol{\mu}_\ss \cdot \S_\pm \, . 
		\end{align}
		where $\boldsymbol{\mu}$ is a vector of chemical potentials. This expression remains valid for $Q \in \mathbb{R}^\ast$. 
		This will play a central role when analyzing the case of a diverging lower bound.
		From \eqref{SM-eq:F_pm}, we interpret the $\boldsymbol{F}_{\pm}$s as coefficients quantifying the response of the stationary chemical potentials to a perturbation of $Q$.

		In what follows, we assume that the matrix $\S_+$ is invertible. We interpret this assumption as the property that 
		the reactions in the autocatalytic network are reactant-independent, in the sense that it not possible to recover the reactants of any reaction by a linear combination of the reactants of other reactions. 
		This means that a given a set of reactants can experience only one reaction together.  
		Under such conditions, $\boldsymbol{F}_+$ is directly related to the derivative of the chemical potential through: $\dQ \boldsymbol{\mu}^\ss =  \boldsymbol{F}_+ \cdot \S_+^{-1}$, yielding the general relation
		\begin{equation}
			\label{SM-eq:Kirchhoff_general}
			\boldsymbol{F}_- = \boldsymbol{F}_+ \cdot \left( \S_+^{-1} \cdot \S_- \right). 
		\end{equation}
		We expect this hypothesis to hold for a wide variety of chemical networks. 
		As an illustration of this, we note that after chemostatting the external species we are often left with a square autocatalytic cycle, in which each reaction consumes only one kind of autocatalytic  species (and possibly many external ones) to form other autocatalytic  species.

		\section*{
			Supplementary Note 4:
			Extension to non-ideal systems}
		
		Our derivation of the structural relations \eqref{SM-eq:Kirchhoff_general} seems to be conditional on the use of mass-action laws, and thus on the assumption of ideality.
		However, we can extend \eqref{SM-eq:Kirchhoff_general} to the case of non-ideal mixtures.
		In such case, one should take include \emph{activity coefficients} $\gamma_\sigma$ in the definition of chemical potentials: $\mu_\sigma = \mu^\circ_\sigma + \, \log (\gamma_\sigma z_\sigma) $.
		Because of thermodynamic consistency, it is  expected that the non-ideal microscopic currents, $j_{\pm \rho}$, obey the de Donder relation
		\begin{equation}
			\label{SM-eq:de_Donder_nonideal}
			\dfrac{j_{+\rho}}{j_{-\rho}} = \exp \left(- \sum_\sigma \mu_\sigma S_\rho^\sigma \right) = e^\mathcal{A}. 
		\end{equation}
		The reaction rates can thus always be written as
		\begin{align}
			\label{SM-eq:Non_ideal_rate}
			j_{+ \rho } = \dfrac{1}{\tau_\rho}  \exp\left( \sum_\sigma \mu_\sigma \, S_{+ \rho}^\sigma  \right), & \quad & 
			j_{- \rho } = \dfrac{1}{\tau_\rho} \exp\left( \sum_\sigma \mu_\sigma \, S_{- \rho}^\sigma  \right),
		\end{align}
		where $\tau_\rho$ is a characteristic timescale of the reaction. \\
		We must recover the ideal kinetic laws when $\gamma_\sigma = 1$, which implies that in such case  
		\begin{equation}\label{SM-eq:constants}
			k_{\pm \rho} = \dfrac{1}{\tau_\rho} \exp \left( \sum_\sigma \mu^\circ_\sigma \, S_{\pm \rho}^\sigma \right). 
		\end{equation}
		Assuming this parametrization to hold in the presence of intermolecular interactions, we note that 
		the derivative of the steady unidirectional flows with respect to $Q$ becomes
		\begin{equation}
			\d_Q j_{\pm \rho}^\ss = \left( \sum_\sigma \d_Q \mu_\sigma^\ss \, S_{\pm \rho}^\sigma \right)  \dfrac{1}{\tau_\rho} \exp \left( \sum_\sigma \mu_\sigma^\ss \, S_{\pm \rho}^\sigma  \right) = F_{\pm \rho} ~ j^\ss_{\pm \rho},
		\end{equation}
		with $F_{\pm \rho} = \d_Q \boldsymbol{\mu}^\ss \cdot \mathbf{S}_{\pm \rho}$, as in \eqref{SM-eq:F_pm}. We thus recover the same structural relations as before, but with a generalized definition of the response coefficients $F_{\pm \rho}$. Note that the expression we use for the kinetic rate constants \eqref{SM-eq:constants} amounts to assuming that the $\tau_\rho$s are constant. Having concentration-dependent time scales (and thus, concentration-dependent rate constants) is a possibility that would not affect  \eqref{SM-eq:de_Donder_nonideal}.  However, this would impact the derivative of the currents $j_{\pm \rho}$ and would strongly modify the structural constraints.

		
		\section*{
			Supplementary Note 5:
			Bound for 3$\times$3 autocatalytic networks}

		Here, we study generic 3$\times$3 autocatalytic networks because this covers all the simplest  autocatalytic cores found in \cite{Blokhuis2020} (except Type~I, which is 2$\times$2). 
		For an arbitrary 3$\times$3 network, $\S_-$ is a 3$\times$3 matrix of positive coefficients of the following form: 
		\begin{equation}
			\S_- = \begin{pmatrix} 0 & \alpha & \beta \\ 
				\gamma & 0 &  \delta\\
				\varepsilon & \zeta & 0 \\
			\end{pmatrix}. 
		\end{equation} 
		Assuming that $\S_+$ is the identity matrix (which is the case for all the autocatalytic cores found in \cite{Blokhuis2020}), the structural constraints Eq.~(27) of the main text can be written as $\boldsymbol{F}_+ \cdot \mathbb{M} = 0$, with
		\begin{equation}
			\mathbb{M} = \begin{pmatrix} e^{\mathcal{A}_1^\ast} & -\alpha & -\beta \\ 
				-\gamma & e^{\mathcal{A}_2^\ast} & -\delta \\
				-\varepsilon  & -\zeta & e^{\mathcal{A}_3^\ast} \\
			\end{pmatrix}.
		\end{equation}	 
		We are looking for non-trivial solutions $\boldsymbol{F}_+ \neq \boldsymbol 0$. As a result, we further enforce $ \det( \mathbb{M} )= 0 $. 
		This yields:
		\begin{equation}
			\label{SM-eq:det_M_3x3}
			e^{\mathcal{A}_1^\ast + \mathcal{A}_2^\ast + \mathcal{A}_3^\ast} = \delta \, \zeta e^{\mathcal{A}_1^\ast } + \beta \, \varepsilon\,  e^{\mathcal{A}_2^\ast} + \alpha \, \gamma\,   e^{\mathcal{A}_3^\ast } + \beta\,  \gamma \, \zeta  + \alpha \, \delta\,  \varepsilon. 
		\end{equation}
		
		Since the elements of $\S$ are natural integers, $\S^{-1}$ is a matrix of rational numbers.
		As a consequence, there exists some \textit{minimal} $n \in \mathbb{N}^\ast $ such that $\hatgX = n \, \gX $ is an integer-valued vector.
		We can thus define $ \widehat{\mathcal{A}} = \sum_\rho \, \widehat{g}_\textsf{X}^\rho \, \mathcal{A}_\rho^\ast = n \, \mathcal{A}^\ast$. If $\gX \geq 0$ (component-wise), so is $\hatgX$, and we can easily find a lower bound for the right-hand side	of \eqref{SM-eq:det_M_3x3}:
		\begin{equation}
			\label{SM-eq:Bound_widehat_A_3x3}
			e^{\widehat{\mathcal{A}}} \geq e^{\mathcal{A}_1^\ast + \mathcal{A}_2^\ast + \mathcal{A}_3^\ast} \geq \delta \, \zeta + \beta \, \varepsilon  + \alpha \, \gamma + \beta\,  \gamma \, \zeta  + \alpha \, \delta \, \varepsilon,
		\end{equation}
		because $g_\textsf{X}^\rho \geq 0 $ implies $ \exp(\mathcal{A}_\rho^\ast) \geq 1 $.
		Consequently, we obtain the following general form for the bound of $\mathcal{A}^\ast$: 
		\begin{equation} 
			\label{SM-eq:general_bound_3x3}
			\mathcal{A}^\ast \geq \dfrac{1}{n} \log \left( \delta \, \zeta + \beta \, \varepsilon  + \alpha\,  \gamma + \beta \, \gamma \, \zeta  + \alpha\,  \delta\,  \varepsilon \right).
		\end{equation}

		Let us assume for example that $g_\textsf{X}^\mathsfit{1} \leq 0$ and 
		$g_\textsf{X}^\mathsfit{2},~g_\textsf{X}^\mathsfit{3} \geq 0$.  Then,
		\begin{align}
			0 < \exp \left( \mathcal{A}_1^\ast \right) \leq 1, & \quad 
			& \exp \left( \mathcal{A}_2^\ast \right) \geq 1, &  \quad &
			\exp \left( \mathcal{A}_3^\ast \right) \geq 1.
		\end{align}
		In such case, the right-hand side of \eqref{SM-eq:det_M_3x3} cannot be lower-bounded directly. 
		Nevertheless, multiplying \eqref{SM-eq:det_M_3x3} by $\exp \left( - 2 \mathcal{A}_1^\ast \right)$ gives:  
		\begin{eqnarray}\nonumber 
			e^{ -\mathcal{A}_1^\ast + \mathcal{A}_2^\ast + \mathcal{A}_3^\ast } &=& \delta\,  \zeta\,  e^{- \mathcal{A}_1^\ast} + \beta \, \varepsilon \, e^{\mathcal{A}_2^\ast - 2 \mathcal{A}_1^\ast} + \alpha \, \gamma \,  e^{\mathcal{A}_3^\ast - 2 \mathcal{A}_1^\ast}\\
			& &  + e^{- 2 \mathcal{A}_1^\ast}\, \left( \beta \, \gamma\,  \zeta  + \alpha \, \delta\,  \varepsilon \right).
		\end{eqnarray}
		Since $\exp \left(  -\mathcal{A}_1^\ast \right) \geq 1$, we recover \eqref{SM-eq:general_bound_3x3}. 
		Generally speaking, the same procedure can be carried out when two arbitrary components of $\gX$ are negative. In this way, we get an equality for $\exp \left( \sum_\rho \text{sgn}(g_\textsf{X}^\rho) \, \mathcal{A}_\rho^\ast \right) $, which reads
		\begin{equation}
			\label{eq:General_3x3}
			e^{\widehat{\mathcal{A}}} \geq \exp \left( \sum_\rho \text{sgn}(g_\textsf{X}^\rho) \, \mathcal{A}_\rho^\ast \right) \geq \delta \, \zeta + \beta \, \varepsilon  + \alpha\,  \gamma + \beta \, \gamma\,  \zeta  + \alpha \, \delta \, \varepsilon.
		\end{equation}
		
		Note that the bound we get from enforcing $\det \left( \mathbb{M} \right) = 0$  may not always be reachable asymptotically for some choice of the kinetic parameters. 
		In fact, a tighter bound can be obtained by incorporating the structural relations (Eq.~(27) of the main text) in \eqref{SM-eq:det_M_3x3},
		as shown below for the derivation of the bound for Type~V networks and for the Hinshelwood cycle.
		
		\section*{
			Supplementary Note 6:
			Extension of the method beyond the 3$\times$3 case}
		
		A relation similar to \eqref{eq:General_3x3} can be derived for an arbitrary 4$\times$4 network. For larger matrices, the method based on the determinant will in general not provide a simple inequality involving the overall affinity. 
		This is because terms of opposite sign can appear in the determinant of $\mathbb{M}$, preventing us from writing an equality between $\exp ( \widehat{\mathcal{A}} ) $ and a sum of positive terms that can easily be bounded. 
		
		Due to the sparsity of $\S_-$ for large networks, the determinant condition still works for a broad class of networks beyond the 3$\times$3 case. 
		In particular if, upon re-arranging the rows and columns,  $\S_-$ can be written:
		\begin{equation}
			\label{SM-eq:General_S_mn_determinant}
			\S_- = \begin{pNiceArray}{ccccw{c}{1.2em} c}[nullify-dots]
				0        & 0      & \Cdots &     &  0     & S_{-N}^1 \\
				S_{-1}^2 &       & \Ddots &     & \Vdots & 0  \\
				\Vdots[shorten=3pt]   & \Ddots & \Ddots[shorten=5pt] &     &        & \Vdots \\ 
				&        &        &     &  0     &      \\
				&        &        &     &  0     & 0  \\
				S_{-1}^N & \Cdots[shorten=5pt] &        &     & S_{-(N-1)}^N & 0 \\
			\end{pNiceArray},
		\end{equation}
		with $S_{\scriptscriptstyle -N}^1 > 0 $, then the  method we presented in Materials and Methods can be generalized to any $N$. 
		The  form \eqref{SM-eq:General_S_mn_determinant} describes what we call a non-intersecting network in the main text.
		Assuming that $\S_-$ satisfies \eqref{SM-eq:General_S_mn_determinant}, we can  expand the first row of $\mathbb{M} =  \text{Diag}(\eA{1}, \cdots, \eA{N}) - \S_-$ as follows:
		\begin{equation}
			\det (\mathbb{M}) = \eA{1} \underbrace{
				\begin{vNiceArray}{ccccc}[nullify-dots]
					\eA{2}    & 0      & \Cdots &          & 0  \\
					-S_{-2}^3 & \Ddots & \Ddots &          & \Vdots \\
					\Vdots    & \Ddots &        &          &    \\
					&        &        &          &  0 \\
					-S_{-2}^N & \Cdots &        & -S_{- (N-1)}^N & \eA{N} \\
				\end{vNiceArray}
			}_{(\text{I})} + (-1)^{N+1} (-S_{-N}^1)  \underbrace{
				\begin{vNiceArray}{ccccc}[nullify-dots]
					-S_{-1}^2 & \eA{2}    & 0      & \Cdots & 0  \\
					-S_{-1}^3 & -S_{-2}^3 & \Ddots & \Ddots & \Vdots  \\
					\Vdots    & \Vdots    & \Ddots &        &  0  \\
					&           &        &        & \eA{N-1}  \\
					-S_{-1}^N & -S_{-2}^N & \Cdots &        & -S_{- (N-1)}^N  \\
				\end{vNiceArray}
			}_{(\text{II})}.
		\end{equation}
		The first term (I) is the determinant of a diagonal matrix, which is given by the product of its diagonal entries.
		The second term (II) can be written as 
		\begin{equation}
			\label{SM-eq:Property_star}
			\text{(II)} = (-1)^{N-1} \sum_i a_i(\S_-) \prod_j (\eA{j})^{\alpha_{ij}},
		\end{equation}
		where the coefficients $a_i(\S_-)$ are positive and depend uniquely on the coefficient in $\S_-$, $\alpha_{ij}$ is either 0 or 1, and $(-1)^{N-1}$ is the sign of (II).
		We can prove property \eqref{SM-eq:Property_star} by induction from determinants having the same shape as (II): a negative lower triangular part, and zeros everywhere else except on the first upper diagonal. Denoting $\Delta_N$ the determinant of an $N \times N$ matrix of this type, we find for 
		the $2 \times 2 $ case  
		\begin{equation}
			\Delta_2 = \begin{vmatrix}
				-S_{-1}^2 & \eA{2} \\[7pt] -S_{-1}^3 & -S_{-2}^3 \\
			\end{vmatrix} = S_{-1}^2 S_{-2}^3 +  S_{-1}^3 \eA{2}.
		\end{equation}
		Assuming \eqref{SM-eq:Property_star} to be true for $N-1$ (with $N\geq 3$) gives  
		\begin{align}
			\Delta_{N} & =  \begin{vNiceArray}{ccccc}[nullify-dots]
				-S_{-1}^2 & \eA{2}    & 0      & \Cdots & 0  \\
				-S_{-1}^3 & -S_{-2}^3 & \Ddots & \Ddots & \Vdots  \\
				\Vdots    & \Vdots    & \Ddots &        &  0  \\
				&           &        &        & \eA{N}  \\
				-S_{-1}^{N+1} & -S_{-2}^{N+1} & \Cdots &        & -S_{- N}^{N+1}  \\
			\end{vNiceArray} \\[10pt]
			& =  \left( -S_{-1}^2 \right)  \cdot 
			\underbrace{
				\begin{vNiceArray}{ccccc}[nullify-dots]
					-S_{-2}^3 & \eA{3}    & 0      & \Cdots & 0  \\
					-S_{-2}^4 & -S_{-2}^4 & \Ddots & \Ddots & \Vdots  \\
					\Vdots    & \Vdots    & \Ddots &        &  0  \\
					&           &        &        & \eA{N}  \\
					-S_{-2}^{N+1} & -S_{-2}^{N+1} & \Cdots &        & -S_{- N}^{N+1}  \\
				\end{vNiceArray}
			}_{ \Delta_{N-1} } - \eA{2} \cdot 
			\underbrace{
				\begin{vNiceArray}{ccccc}[nullify-dots]
					-S_{-1}^3 & \eA{3}    & 0      & \Cdots & 0  \\
					-S_{-1}^4 & -S_{-2}^4 & \Ddots & \Ddots & \Vdots  \\
					\Vdots    & \Vdots    & \Ddots &        &  0  \\
					&           &        &        & \eA{N}  \\
					-S_{-1}^{N+1} & -S_{-2}^{N+1} & \Cdots &        & -S_{- N}^{N+1}  \\
				\end{vNiceArray}
			}_{\Delta^\prime_{N-1}} \label{SM-eq:Induction_star}. 
		\end{align}
		The two determinants $\Delta_{N-1}$ and $\Delta^\prime_{N-1}$ obey \eqref{SM-eq:Property_star}, hence factorizing the $(-1)$ in \eqref{SM-eq:Induction_star} yields: $\mathrm{sgn}\left( \Delta_N \right) = (-1)^{N}$.
		We are then left with a sum of products of coefficients in $\S_-$ (which are positive) and of the $\eA{i}$s.
		In addition, each $\eA{i}$ appear, at most once in each term of the sum because they are present at most once in the determinants $\Delta_{N-1}$ and $\Delta^\prime_{N-1}$. 
		This completes the proof of the property \eqref{SM-eq:Property_star} \\
		
		Enforcing $\det (\mathbb{M}) = 0$ yields:
		\begin{equation}
			\exp \left( \sum_\rho \mathcal{A}_\rho^\ast \right)  = S_{-N}^1 \sum_i a_i \left( \S_- \right)  \prod_j (\eA{j})^{\alpha_{ij}}.
		\end{equation}
		This equation provides us with a bound on $\widehat{\mathcal{A}} = n \,\mathcal{A}^\ast$ (where again, $n$ is the lowest positive integer that makes the product $n \, \gX$  an integer-valued vector):
		\begin{equation}
			e^{n \mathcal{A}^\ast}  \geq S_{-N}^1 \sum_i a_i \left( \S_- \right) .
		\end{equation}

		\section*{
			Supplementary Note 7:
			Case of networks with only one branched reaction} 
		
		From \eqref{SM-eq:General_S_mn_determinant}, we can derive a simple expression for the lower bound of the affinity for a network containing only one branching reaction with $\mathfrak{p} \geq 2$ products.
		In that case, we can write
		\begin{equation}
			\S_- = \begin{pNiceMatrix}[nullify-dots]
				0 & 0 & \Cdots &  & 0 & 1 \\[0.5em]
				S_{-1}^2 & \Ddots[shorten=5pt] & \Ddots[shorten=2pt] &  &  & 0 \\ 
				S_{-1}^3 & 1 & &  &   &  \Vdots \\ 
				\Vdots & 0 & \Ddots &   &   &  \\ 
				& \Vdots &  \Ddots &   &   &  0 \\
				S_{-1}^N & 0 & \Cdots  & 0 & 1 & 0 \\
			\end{pNiceMatrix},
		\end{equation}
		with $ \sum_{i=2}^N S_{-1}^i = \mathfrak{p}$. 
		Upon expanding the first row of $\mathbb{M}$, we now find 
		\begin{equation}
			\det \left( \mathbb{M} \right)  = \exp \left( \sum_\rho \mathcal{A_\rho^\ast} \right)  + (-1)^{N+2} \underbrace{\begin{vNiceMatrix}[nullify-dots]
					- S_{-1}^2 & \eA{2} & 0 & \Cdots &  & 0 \\ 
					- S_{-1}^3 & -1 & \Ddots & \Ddots &  & \Vdots \\ 
					\Vdots & 0 & \Ddots[shorten=5pt] &  &  &  \\
					& \Vdots  & \Ddots &  &  & 0 \\
					&  &   &  &  & \eA{N-1} \\
					- S_{-1}^N & 0 & \Cdots &  & 0 & -1  \\
			\end{vNiceMatrix}}_{\Delta_{N}^{2} }.
		\end{equation}
		Moreover, we have the recurrence relation $\Delta^i_N = (-1)^{N-i+1} S^i_{-1} - \eA{i} \Delta_{N}^{i+1} $ for $ 2 \leq i \leq N$, with:
		\begin{align}
			\Delta^i_N = \begin{vNiceMatrix}
				- S_{-1}^i & \eA{i} & 0 & \Cdots & 0 \\ 
				- S_{-1}^{i+1} & -1 & \Ddots & \Ddots & \Vdots   \\ 
				\Vdots & 0 & \Ddots &  &  0  \\
				& \Vdots &  \Ddots &  &  \eA{N-1}  \\
				- S_{-1}^N & 0 & \Cdots  & 0 & -1  \\
			\end{vNiceMatrix}, & \quad & \Delta_N^N = -S^N_{-1},
		\end{align}
		giving $\Delta_N^2 = (-1)^{N-1} \sum_{i=2}^N S_{-1}^i  \prod_{j=2}^{i-1} \eA{j}$ (note that the latter is in accordance with \eqref{SM-eq:Property_star}). 
		We now find that  
		\begin{equation}
			\label{SM-eq:Bound_one_fork}
			e^{n \mathcal{A}^\ast} \geq \sum_{i=2}^N S^i_{-1} = \mathfrak{p}.
		\end{equation}
		Throughout all this section we assumed $\S_+ = \text{Id}$. If this is not the case, the above results can be generalized by replacing $\S_-$ by $(\S_+^{-1} \cdot \S_-)$, provided $(\S_+^{-1} \cdot \S_- )$ satisfies \eqref{SM-eq:General_S_mn_determinant}. 
		That will be true if, for example, $\S^+$ is a diagonal matrix and $\S_-$ satisfies \eqref{SM-eq:General_S_mn_determinant}.  
		
		\section*{
			Supplementary Note 8:
			The Generalized Type~I}
		
		The generalization of Type~I networks constitutes a significant class of systems.  In this generalization,  $\alpha$ species \textsf{X} pass through an arbitrary number of linear intermediates of $Y$ species,  giving back $\beta \,$\textsf{X}, $(\beta > \alpha)$:
		\begin{equation}
			\label{SM-eq:Generalized_Type_I_network}
			\ce{\alpha \, \textsf{X} <--> $\textsf{Y}_1$ <--> $\cdots$ <--> $\textsf{Y}_{n}$ <--> \beta \, \textsf{X}}. 
		\end{equation}
		For that network: 
		\begin{align}
			\label{SM-eq:Topology_Generalized_Type_I}
			\S_+ = \text{Diag} (\alpha, 1, \cdots, 1),  & \quad & 
			\S_- = \begin{pNiceMatrix}[nullify-dots]
				0 & 0 & \Cdots & 0 & \beta \\
				1 & \Ddots &  \Ddots &  &  0 \\
				0 & \Ddots &  &  & \Vdots \\
				\Vdots & \Ddots &  &  & 0	\\
				0 & \Cdots & 0 & 1 & 0	
			\end{pNiceMatrix},
		\end{align}
		and the determinant of $\S = \S_- - \S_+$ is $(-1)^{n} (\beta - \alpha)$. Thus, $\S$ is invertible if and only if $\beta \neq \alpha$, namely, only if \eqref{SM-eq:Generalized_Type_I_network} is indeed an autocatalytic network. 
		If \textsf{X} is chemostatted, its elementary mode of production is \emph{i.e.} the first column of $\S^{-1}$, which is given by  $\boldsymbol{g}_\textsf{X} = (\beta - \alpha)^{-1}\, \boldsymbol{1} > 0$, where $\boldsymbol{1}$ is a vector of ones. 
		Furthermore, $(\S_+^{-1}\cdot\S_-)$ has a shape of type \eqref{SM-eq:General_S_mn_determinant}, and enforcing the determinant of ${\mathbb{M} = \text{Diag}(e^{\mathcal{A}_1^\ast}, \cdots, e^{\mathcal{A}_{n+1}^\ast} ) - ( \S_+^{-1} \cdot \S_- )}$ to vanish yields
		\begin{equation}
			\label{SM-eq:Bound_Generalized_Type_I}
			e^{(\beta - \alpha)\mathcal{A}^\ast} = \dfrac{\beta}{\alpha}.
		\end{equation}
		In particular, consider $\widehat{\boldsymbol{g}}_\textsf{X} = (\beta - \alpha) \, \boldsymbol{g}_\textsf{X} = \boldsymbol{1}$, which is the pathway producing $\beta$ \textsf{X}s starting from $\alpha$ \textsf{X}s. The overall affinity at $Q^\ast$ is always $ \log (\beta/\alpha) $ or, equivalently, $Q^\ast = (\beta/\alpha) K $.
		When $\alpha = 1$ and $\beta = 2$ one recovers the result derived in the main text for the Type~I core. 
		
		Going beyond chemical networks, Type~I and its generalized version can also be seen as a  coarse-grained representation of more general networks and the bounds will be unchanged thanks to a property of network reduction explained in the note 10 below. 
		The overall autocatalytic process can be  described by the balance equation \ce{$\alpha$ \textsf{X} + $\mathrm{(\diamond)}$ -> $\beta$ \textsf{X} + $\mathrm{(\diamond)}$}, which we called a seed-dependent mode of production in the main text. 
		We note that if one adds non-zeros terms to lower diagonal of $\S_-$ in such systems, the property \eqref{SM-eq:General_S_mn_determinant} is still verified and the lower bound either stays the same or increases. 
		This shows our conjecture concerning the seed-dependent modes of production for non-intersecting networks.

		\NiceMatrixOptions{
			code-for-last-row = \color{black!70}\scriptstyle,
			code-for-first-col = \color{black!70}\scriptstyle,
		}

		\section*{
			Supplementary Note 9:
			The Generalized Type~II}
		
		Looking at Table 1 in Methods section of the main text, we notice that the Type~II network exhibits a singular behavior when species \textsf{A} is chemostatted. 
		In that case, the affinity also saturates at $\log 2$, just like for Type~I networks.
		This result is obtained by enforcing $\det \left( \mathbb{M} \right) = 0$, where 
		\begin{equation}
			\mathbb{M} = \begin{pmatrix}
				\eA{1} & 0 & -1 \\ -1 & \eA{2} & -1 \\ 0 & -1 & \eA{3}
			\end{pmatrix}, 
		\end{equation}
		with $\eA{1} = 1$ because $g_\textsf{X}^\mathsfit{1} = 0$.
		Furthermore, $g_\textsf{X}^\mathsfit{1} = 0$ implies that this mode of production is \emph{seed-independent}, as defined in the main text. 
		Type II networks represent, in fact, the simplest way of producing \textsf{A} without the need to have \textsf{A} initially. 
		
		This observation motivates the introduction of a generalized Type~II network: 
		\begin{equation}
			\label{SM-eq:Generalized_Type_II_network}
			\ce{ \alpha \, \textsf{X} 
				<-->[$\mathsfit{1}$] 
				$\textsf{Y}_{1}$
				<-->[$\mathsfit{2}$]
				$\cdots$ 
				<-->[$\mathsfit{n}$]
				$\textsf{Y}_{n}$ <-->[$\mathsfit{n+1}$] 
				$\textsf{Y}_1$ + \beta \, \textsf{X},
			}
		\end{equation}
		which verifies,
		\begin{align}
			\label{SM-eq:S_Generalized_Type_II}
			\S_- = \begin{pNiceMatrix}[nullify-dots, last-row, first-col]
				\mathsfit{Y}_n & 0 & 0 & \Cdots &  & 0 & 1 \\
				\mathsfit{X} & \beta & 0 & \Ddots &  &  & 0 \\ 
				\mathsfit{Y}_1 & 1 & 1 & \Ddots & &  & \Vdots \\ 
				\mathsfit{Y}_2 & 0 & 0 & \Ddots & &  &  \\
				& \Vdots & \Vdots & \Ddots & & & 0\\
				\mathsfit{Y}_{n-1}  & 0 &  0 & \Cdots & 0 & 1 & 0 \\
				& \mathsfit{n+1} & \mathsfit{1} & \Cdots & & \mathsfit{n-1} & \mathsfit{n}
			\end{pNiceMatrix},
			& \quad & 
			\S_+ = \text{Diag} (1, \alpha, 1, \cdots, 1).
		\end{align}
		In addition, the rescaled mode of production of species \textsf{X} can be easily constructed,  $\widehat{\boldsymbol{g}}_\textsf{X} = \left(0, 1, 1, \cdots, 1 \right)^\top$. 
		Indeed, the latter corresponds to the overall equation:
		\begin{equation}
			\label{SM-eq:Generalized_Type_II}
			\ce{
				$\textsf{Y}_{1}$ +  $\cdots$ +  $\textsf{Y}_{n}$ 
				-> 
				$\textsf{Y}_{1}$ +  $\cdots$ +  $\textsf{Y}_{n}$ +  \beta\,  \textsf{X}
			}.
		\end{equation}
		This overall equation is seed-independent because species \textsf{X} is not needed as a reactant for its own production, it is of the form of Eq.~(33) in the Main Text in which, here,  $\mathrm{(\diamond)} \equiv \textsf{Y}_{1}$ +  $\cdots$ +  $\textsf{Y}_{n}$. 
		The generalized Type~II network is the simplest topology that admits a seed-independent overall equation following Eq.~(33). 
		The matrix $\mathbb{M}$ reads  
		\begin{equation}
			\mathbb{M} = \begin{pNiceMatrix}[nullify-dots]
				\eA{n+1} & 0 & \Cdots &  & 0 & -1 \\
				-\beta/\alpha & \eA{1} & \Ddots &  &  & 0 \\ 
				-1 & -1 & \Ddots & &  & \Vdots \\ 
				0 & 0 & \Ddots & &  &  \\
				\Vdots & \Vdots & \Ddots & & & 0\\
				0 &  0 & \Cdots & 0 & -1 & \eA{n}\\
			\end{pNiceMatrix}.
		\end{equation}
		Expanding along the first row of $\mathbb{M}$ gives for the determinant $\det \left( \mathbb{M} \right) = e^{\mathcal{A}^\ast} - (\beta/\alpha + \eA{1})$. 
		The first reaction should be at equilibrium as its associated entry in the elementary mode vanishes hence, $\mathcal{A}^\ast_1 = 0$. 
		This yields an analogue of \eqref{SM-eq:Bound_Generalized_Type_I} for seed-independent production: 
		\begin{equation}
			\label{SM-eq:Bound_Generalized_Type_II}
			e^{\mathcal{A}^\ast} = \dfrac{\beta}{\alpha} + 1. 
		\end{equation}
		Again, adding terms in the lower diagonal of $\S_-$ in \eqref{SM-eq:S_Generalized_Type_II} necessarily increases the bound \eqref{SM-eq:Bound_Generalized_Type_II}, which supports the conjecture made in the main text. 
		
		Finally, in a generalized Type~II with no intermediate species ($n=0$), \ce{$\varnothing$ <--> \beta\,  \textsf{X}} is the only microscopic reaction in the network \emph{and} the overall reaction associated to the production of  $\textsf{X}$. 
		In that case, the method based on the determinant does not work since $\S_+ = 0$ is not invertible. 
		However, the analysis becomes trivial since the macroscopic current identifies to the only microscopic current which is strictly decreasing with respect to the concentration of $\textsf{X}$, and attains its maximum at zero. 
		This does not contradict the proposed bound as this yields a diverging affinity.
		
		\section*{
			Supplementary Note 10:
			Network reduction}
		
		\NiceMatrixOptions
		{nullify-dots, code-for-first-col = \color{black!70}\scriptstyle, code-for-last-row=\color{black!70}\scriptstyle}
		
		Here, we show that the overall affinity of a chemical network in the regime of optimal flux considered in this article can be calculated by evaluating it on a smaller network, provided certain conditions are met. By definition, the reduced network contains only \emph{essential species}, and is obtained from the initial network by a coarse-graining procedure, by which non-essential 
		\emph{intermediate species} are removed from the description. For simplicity, let us assume that these intermediate species are present only in linear segments connecting two essential species, which we call \emph{branches}. 
		Before considering the general case, we illustrate below the method using an example with a Type~II network.

		\paragraph{Example for the reduction of a Type~II-like network}
		
		Let us consider the following network, which consists of three branches $\mathcal{B}_1, \mathcal{B}_2$ and 
		$\mathcal{B}_3$:
		
		\begin{minipage}{0.25\textwidth}
			\centering
			\includegraphics{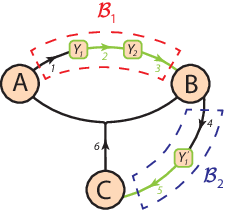}
		\end{minipage}
		\begin{minipage}{0.75\textwidth}
			\begin{align}
				\label{SM-eq:Augmented_Type_II}
				\quad \begin{cases}
					\mathcal{B}_1: ~ \ce{\textsf{A} <-->[$\mathsfit{1}$] $\mathsfit{Y}_1$ <-->[$\mathsfit{2}$] $\mathsfit{Y}_2$ <-->[$\mathsfit{3}$] \textsf{B} } \\[0.5em]
					\mathcal{B}_2: ~ \ce{\textsf{B} <-->[$\mathsfit{4}$] $\mathsfit{Y}^\prime_1$ <-->[$\mathsfit{5}$] \textsf{C} } \\[0.5em]
					\mathcal{B}_3: 
					~ \ce{\textsf{C} <-->[$\mathsfit{6}$]  \textsf{A} + \textsf{B} }
				\end{cases}, & \quad & 
				\S = ~ 
				\begin{pNiceMatrix}[last-row, first-col]
					\textsf{A} &  -1 & 0  & 0 & 0 & 0 & 1 \\
					\mathsfit{Y}_1 & 1 & -1 & 0 & 0 & 0 & 0 \\ 
					\mathsfit{Y}_2 & 0 &  1 & -1 & 0 & 0 & 0 \\
					\textsf{B} &   0 & 0  & 1 & -1 & 0 & 1 \\
					\mathsfit{Y}^\prime_1 & 0 & 0 & 0 & 1 & -1 & 0 \\ 
					\textsf{C} &   0 & 0  & 0 & 0 & 1 & -1 \\
					& \mathsfit{1} & \mathsfit{2} & \mathsfit{3} & \mathsfit{4} & \mathsfit{5} & \mathsfit{6}
				\end{pNiceMatrix}.  
			\end{align}
		\end{minipage}\\
		The only branched reaction is $\mathsfit{6}$ and, in this network, the essential species are
		$\left\lbrace \textsf{A}, \textsf{B}, \textsf{C} \right\rbrace $, while  the intermediate species are $\left\lbrace  \mathsfit{Y}_1, \mathsfit{Y}_2, \mathsfit{Y}^\prime_1  \right\rbrace$. 
		The essential network is:
		
		{\begin{minipage}{0.2\textwidth}
				\centering
				\includegraphics{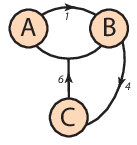}
			\end{minipage}
			\begin{minipage}{0.75\textwidth}
				\begin{align}
					\label{SM-eq:Type_II_reduction}
					\begin{cases}
						\ce{\textsf{A} <-->[$\mathsfit{1}$] \textsf{B} } \\
						\ce{\textsf{B} <-->[$\mathsfit{4}$] \textsf{C} } \\
						\ce{\textsf{C} <-->[$\mathsfit{6}$]  \textsf{A} + \textsf{B} }
					\end{cases}, & \quad & 
					\S^\text{ess} = ~ 
					\begin{pNiceMatrix}[last-row, first-col]
						\textsf{A} &  -1 &  0 & 1 \\
						\textsf{B} &   1 &  -1 & 1 \\
						\textsf{C} &   0 & 1 & -1 \\
						& \mathsfit{1}  & \mathsfit{4} & \mathsfit{6}
					\end{pNiceMatrix}, & \quad &
					\left( \S^\text{ess}\right)^{-1}  = ~ 
					\begin{pNiceMatrix}[last-row, first-col]  
						\mathsfit{1} & 0 & 1 & 1 \\ 
						\mathsfit{2} & 1 & 1 & 2 \\ 
						\mathsfit{3} & 1 & 1 & 1 \\ 
						& \bm{\mathsfit{g}}_\textsf{A}^\text{ess}   & \bm{\mathsfit{g}}_\textsf{B}^\text{ess} & \bm{\mathsfit{g}}_\textsf{C}^\text{ess} \\
					\end{pNiceMatrix},
				\end{align}
		\end{minipage}}
		which is a Type~II core. 
		When the chemical network operates in the regime of optimal flux, the following structural relations hold:
		\begin{align}
			F_{-1} = F_{+2}, & \quad &
			F_{-2} = F_{+3}, & \quad &
			F_{-3} = F_{+4}, & \quad &
			F_{-4} = F_{+5}, & \quad &
			F_{-5} = F_{+6}, & \quad &
			F_{-6} = F_{+1} + F_{+4}.
		\end{align}
		From these relations, we can see that the affinity of each branch of the network takes the same value in the original and in the reduced networks, where they are denoted $\mathcal{A}_i^\text{ess}$ for reaction $i$:
		\begin{gather}
			e^{\mathcal{A} \left( \mathcal{B}_1 \right) } = 
			\dfrac{F_{-1} F_{-2} F_{-3}}{F_{+1} F_{+2} F_{+3}} = \dfrac{F_{-3}}{F_{+1}} = \dfrac{F_{-1}^\text{ess}}{F_{+1}^\text{ess}} = e^{\mathcal{A}_1^\text{ess}} \label{SM-eq:Ex_Reduction_Branch_1}, \\[1em]
			e^{\mathcal{A} \left( \mathcal{B}_2 \right) } = 
			\dfrac{F_{-4} F_{-5}}{F_{+4} F_{+5}} = \dfrac{F_{-5}}{F_{+4}} = \dfrac{F_{-4}^\text{ess}}{F_{+4}^\text{ess}} = e^{\mathcal{A}_4^\text{ess}} \label{SM-eq:Ex_Reduction_Branch_12}, \\[1em]
			e^{\mathcal{A} \left( \mathcal{B}_3 \right) } = e^{\mathcal{A}_6^\text{ess}} \label{SM-eq:Ex_Reduction_Branch_3}.
		\end{gather}
		
		Let us assume for instance that species \textsf{A} is chemostatted. Its mode of production in the essential network is the following set of reactions : 
		\begin{equation}
			\boldsymbol{g}_\textsf{A}^\text{ess}:~ \ce{\textsf{B} ->[\mathsfit{4}] \textsf{C} ->[\mathsfit{6}] \textsf{B} + \textsf{A}},
		\end{equation}
		which means that $\boldsymbol{g}_\textsf{A}^\text{ess} = \left( 0, 1, 1 \right)^\top $.
		Similarly, the corresponding vector for the original network is 
		\begin{align}
			\boldsymbol{g}_\textsf{A} = \begin{pNiceMatrix}[last-row]
				0 & 0 & 0 & 1 & 1 & 1 \\ 
				\mathsfit{1} & \mathsfit{2} & \mathsfit{3} & \mathsfit{4} & \mathsfit{5} & \mathsfit{6}
			\end{pNiceMatrix}^\top. 
		\end{align} 
		Note that intermediate species that come between two essential species are described either with a zero entry in that vector if the branch is not present in the mode (for reactions 2 and 3), or with a one if it is present (for reaction 5).
		Thanks to this property, it is straightforward to relate components of this reaction vector in the original and in the reduced networks.

		Then, the initial sum over reactions in the global affinity can  be reduced to a sum over affinities of the branches, which as we explained above take the same value in the original and in the reduced networks. 
		It follows from this that the overall affinity in the complete network reduces to  its value in the essential network when evaluated in the regime of optimal flux : 
		\begin{eqnarray}
			e^{\mathcal{A}^\ast} &=& \exp \left(  \sum_{\rho} g_\textsf{A}^\rho \, \mathcal{A}_\rho \right)  
			=
			\exp \left(  \mathcal{A}_4+ \mathcal{A}_5 + \mathcal{A}_6 \right), \\
			&=&
			\exp \left( \mathcal{A} \left( \mathcal{B}_2 \right) + \mathcal{A} \left( \mathcal{B}_3 \right)  \right)
			\\
			&=& e^{\mathcal{A}^\text{ess}}.
		\end{eqnarray}
		The same property can be shown to hold when other species are chemostatted.
		
		\paragraph{General case}
		
\begin{figure}[h]
			\centering
			\includegraphics{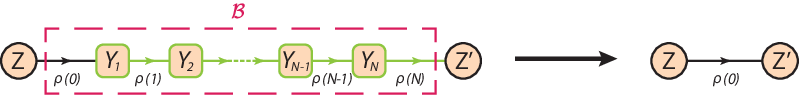}
			\caption{\label{fig:Reduction_branch}
				The branch $\mathcal{B}$ between two essential species $\textsf{Z}$ and $\textsf{Z}'$ is the sub-network formed by the succession of $N$ \emph{intermediate species} (represented in green), connected together by unimolecular non-essential reactions (represented in green).
				The reduction of $\mathcal{B}$ consists in removing all the intermediates, keeping only the essential species. 
			}
		\end{figure}

		The general proof follows exactly the same steps as in the example above. 
		We consider a branch $\mathcal{B}$, which is defined as the sub-network of $N$ \emph{intermediate species} between two \emph{essential species} $\textsf{Z}$ and $\textsf{Z}'$ (see Fig.~\ref{fig:Reduction_branch}). 
		
		The first step is to show that the affinities take the same value for that branch in the original and in the reduced networks, when evaluated in the regime of optimal flux. 
		This follows from the expression of the $\mathcal{A}_\rho^\ast$s as a function of the the $F_{\pm \rho}$s (Eq.~(43) of the main text) and the structural relations in a linear segment:
		\begin{equation}	
			\mathcal{A}(\mathcal{B})   = \log \left( 
			\prod_{i=0}^{N} \dfrac{F_{-\rho(i)}}{F_{+\rho(i)}} 
			\right) 
			= \log \left( \dfrac{ F_{{-\rho}(N)}}{ F_{{+\rho(0) }}} 
			\right)
			= \log \left( \dfrac{ F_{{-\rho}(0)}^\text{ess} }{ F_{{+\rho(0) }}^\text{ess}} 
			\right)
			= \mathcal{A}^{\text{ess}}_{\rho (0)}.
		\end{equation}
		From now on we assume that the chemostatted species \textsf{X} is an essential species that is still present in the network after  reduction.
		The restriction of $\gX$ to the reactions in $\mathcal{B}$ is a constant vector whose values are given by $g_\textsf{X}^{\rho (0)}$. 
		Hence, for all reactions $\rho (i) \in \mathcal{B}$, the corresponding entry in $\gX$ will be $g_\textsf{X}^{\rho (i)} = g_\textsf{X}^{\rho (0)}$. 
		Moreover, the entry $g_\textsf{X}^{\rho (0)}$ is preserved after the reduction: $g_\textsf{X}^{\rho (0)} = (g_\textsf{X}^\text{ess})^{\rho (0)}$. \\
		As a result, here again, the initial sum over reactions in the global affinity can then be reduced to a sum over affinities of the branches, which take the same value in the original and in the reduced networks. 
		It follows from this that the overall affinity in the complete network equals the value it takes in the reduced network, namely $\mathcal{A}^\text{ess}$, irrespective of which essential species is chemostatted: 
		\begin{eqnarray}
			e^{\mathcal{A}^\ast} &=& \exp \left(  \sum_{\rho} g_\textsf{X}^\rho \, \mathcal{A}_\rho \right)  
			=
			\exp \left( \sum_{\mathcal{B}} \sum_{\rho (i) \in \mathcal{B}} g_\textsf{X}^{\rho (i) }\, \mathcal{A}_{\rho (i)} \right), \\
			&=&
			\exp \left(  \sum_{\mathcal{B}} g_\textsf{X}^{\rho(0)}
			\sum_{\rho (i) \in \mathcal{B}} 
			\mathcal{A}_{\rho (i)} \right) 
			= 
			\exp \left( \sum_{\mathcal{B}}   g_\textsf{X}^{\rho(0)}\mathcal{A}(\mathcal{B}) \right) , 
			\\
			&=& e^{\mathcal{A}^\text{ess}}.
		\end{eqnarray}

		\section*{
			Supplementary Note 11:
			Optimization Problem}
		
		The affinity at the optimal current $\mathcal{A}^*$ defines the objective function that we should minimize: 
		\begin{equation}
			\mathcal{A}^* (\boldsymbol{F}_+, \boldsymbol{F}_-) 
			= 
			\sum_{i = 1}^N g_\textsf{X}^i \, \log \left( \dfrac{F_{-i}}{F_{+i}} \right) = 
			\left. \left. \sum_{i = 1}^N g_\textsf{X}^i \right\lbrace \log F_{-i} - \log F_{+i} \right\rbrace. 
		\end{equation}
		The Hessian matrix of $\mathcal{A}^*$ is diagonal:
		\begin{equation}
			\mathbb{H}_{\mathcal{A}} =
			\begin{pmatrix}
				\mathbb{H}_{\mathcal{A}}^- &  & O \\[12pt]
				O &  & \mathbb{H}_{\mathcal{A}}^+ \\
			\end{pmatrix}~, \text{~~ where ~~} \mathbb{H}_{\mathcal{A}}^\pm = \mathrm{Diag} \left[ \left\lbrace \pm \dfrac{g_\textsf{X}^i}{F_{\mp i}^2}  \right\rbrace_{1 \leq i \leq N} \right]
		\end{equation}
		and has positive eigenvalues (in $\mathbb{H}_{\mathcal{A}}^+$) and negative eigenvalues (in $\mathbb{H}_{\mathcal{A}}^-$). 
		Consequently, the affinity $\mathcal{A}^*$ is neither a concave nor a convex function of the $F_\pm$s. 
		We can incorporate the structural constraints \eqref{SM-eq:Kirchhoff_general} into the affinity, denoting $\mathbb{T} = \S_+^{-1} \cdot \S_- = \left\lbrace T^i_j \right\rbrace $, we have:
		\begin{equation*}
			\mathcal{A}^\ast = \sum_{i=1}^N  g_\textsf{X}^i \log \left( \dfrac{\sum_{j=1}^N F_{+ j} \, T^j_i }{F_{+i}} \right).
		\end{equation*}
		Consequently, the affinity can be re-written as a function of the new set of $N^2$ variables $\Gamma^j_i = F_{+j}/F_{+i}$:
		\begin{equation}
			\label{SM-eq:Hessian_Psi_def}
			\mathcal{A}^* (\boldsymbol{F}_+, \boldsymbol{F}_-) = \varPsi \left( \left\lbrace \Gamma^j_i \right\rbrace_{1 \leq i, \, j \leq N} \right) =
			\sum_{i=1}^N  g_\textsf{X}^i \log \left( \sum_{j=1}^N T^j_i \, \Gamma^j_i \right).
		\end{equation} 
		The Hessian matrix of $\varPsi$ is a $N^2 \times N^2$ block-diagonal matrix:
		\begin{equation}
			\mathbb{H}_{\varPsi} = 
			\begin{pNiceMatrix}[nullify-dots]
				\mathbb{H}_{\varPsi}^{(1)} &  &  &  &  &  \\
				&  & & &  O &  \\ 
				&  & \Ddots &  &  &  \\ 
				&  &  &  &  &  \\
				& O  &  &  &  &  \\
				&  &  &  &  & \mathbb{H}_{\varPsi}^{(N)}\\
			\end{pNiceMatrix}, 
		\end{equation}
		where $\mathbb{H}_{\varPsi}^{(i)}$ is a $N \times N$ matrix whose components are: 
		\begin{equation}
			\left( \mathbb{H}_{\varPsi}^{(i)} \right)^k_l = 
			\partial^2_{\scriptsize \Gamma^k_i, \, \Gamma^l_i} \varPsi =
			- \dfrac{g_\textsf{X}^i}{\left( \sum_{j=1}^N T^j_i \, \Gamma^j_i \right)^2} \, T_i^k \, T_i^l.
		\end{equation}
		Now, considering $ \boldsymbol{v}^{(i)} = \left( v^{(i)}_1 , \cdots, v^{(i)}_N \right)^\top $,  we have: 
		\begin{equation}
			{\boldsymbol{v}^{(i)}}^\top \cdot \mathbb{H}_{\varPsi}^{(i)} \cdot \boldsymbol{v}^{(i)} = 
			- \dfrac{g_\textsf{X}^i}{\left( \sum_{j=1}^N T^j_i \, \Gamma^j_i \right)^2}  \, 
			\left( \sum_{j=1}^N T_i^j \, v^{(i)}_j \right)^2
		\end{equation}
		And, for $ \boldsymbol{v} = \left( {\boldsymbol{v}^{(1)}}^\top, \cdots,   {\boldsymbol{v}^{(N)}}^\top \right)^\top $, we have:
		\begin{equation}
			\label{SM-eq:Hessian_Psi}
			\boldsymbol{v}^\top \cdot \mathbb{H}_{\varPsi} \cdot \boldsymbol{v} = 
			\sum_{i=1}^N {\boldsymbol{v}^{(i)}}^\top \cdot \mathbb{H}_{\varPsi}^{(i)} \cdot \boldsymbol{v}^{(i)} =
			- \sum_{i=1}^N g_\textsf{X}^i  \left( \dfrac{  \sum_{j=1}^N T_i^j \, v^{(i)}_j  }{\sum_{j=1}^N T_i^j \, \Gamma^j_i} \right)^2. 
		\end{equation}
		In the general case where $\gX$ has negative and positive values, the Hessian matrix is neither positive nor negative (because the sign of \eqref{SM-eq:Hessian_Psi} can vary with the components of $\boldsymbol{v}$). 
		However, considering the case where $\gX \geq 0$ (component-wise) which appears quite often, the sign of \eqref{SM-eq:Hessian_Psi} is always negative.
		As a consequence, the Hessian of $\varPsi$ is negative (semi-definite), which makes $\varPsi$ a concave function. 
		Hence, when $\gX \geq 0$, the optimization problem associated to $\mathcal{A}^*$ is turned into the minimization of a concave function with constraints:
		\begin{gather}
			\underset{\Gamma}{\mathrm{min}} \left[ \sum_{i=1}^N  g_\textsf{X}^i \log \left( \sum_{j=1}^N T_i^j \, \Gamma^j_i \right) \right] ; \\[12pt]
			\forall 1 \leq i \leq N, ~~ \begin{cases}
				g_\textsf{X}^i > 0 \Rightarrow \sum_j T_i^j \, \Gamma^j_i > 1 \\[10pt]
				g_\textsf{X}^i = 0 \Rightarrow \sum_j T_i^j \, \Gamma^j_i = 1
			\end{cases} \label{SM-eq:Hessian_Additional_Cons} \\[12pt]
			\forall 1 \leq i, \, j, \, k \leq N, ~ \Gamma^i_j \, \Gamma^j_k = \Gamma^i_k \text{~~ and ~~} \Gamma_i^i = 1 \label{SM-eq:Hessian_Var_Cons} .
		\end{gather}
		Where the constraints \eqref{SM-eq:Hessian_Additional_Cons} translate the \emph{additional constraints} Eq.~(44) in the Methods section of the main text. 
		While the last set of constraints \eqref{SM-eq:Hessian_Var_Cons} arise from our choice of variables $\Gamma$. 
		It is worth noted that the unconstrained minimization of a concave function that is not convergent yields infinity, but here the constraints define a feasible domain in the space and the minimum of $\varPsi$ is attained for a set of value $\Gamma$ that lies at the edge of the feasible zone. 
		
		\paragraph{Example of the Generalized Type I}
		In case of the Generalized Type I with one linear intermediate:
		\begin{align}
			\begin{cases}
				\ce{\alpha \, \textsf{A} <--> \textsf{B} } \\
				\ce{\textsf{B} <--> \beta \, \textsf{A}}
			\end{cases}, & \quad & 
			\S = \begin{pmatrix} -\alpha & \beta \\ 1 & -1 \end{pmatrix}, & \quad &
			\left( \beta - \alpha\right) \S^{-1} = \begin{pmatrix} 1 & \beta \\ 1 & \alpha \end{pmatrix}, & \quad & 
			\begin{cases}
				F_{-1} = F_{+2} \\
				\alpha \, F_{-2} = \beta \, F_{+1}
			\end{cases} ;
		\end{align}
		where we multiplied $\S^{-1}$ by $\beta - \alpha$ in order to have integer-valued mode of production $\gX$, for \textsf{X} = \textsf{A} or \textsf{B}. 
		The function $\varPsi$ \eqref{SM-eq:Hessian_Psi_def} that incorporates the structural constraints reads, 
		\begin{equation}
			\varPsi \left( \Gamma \right) = g_\textsf{X}^1 \log  \Gamma^2_1  + g_\textsf{X}^2 \log \left( \dfrac{\beta}{\alpha} \Gamma^1_2 \right), 
		\end{equation} 
		with the following set of constraints:
		\begin{gather}
			\Gamma^2_1 > 1 ~~ \text{and} ~~ \dfrac{\beta}{\alpha} \, \Gamma^1_2 > 1, \\[0.5em] 
			\Gamma^2_1 \, \Gamma^1_2 = 1.
		\end{gather}
		These constraints imply $1 < \Gamma^2_1 < \dfrac{\beta}{\alpha}$ thus:
		\begin{equation}
			\varPsi = 
			g_\textsf{X}^2 \log \dfrac{\beta}{\alpha} + \left( g_\textsf{X}^1 - g_\textsf{X}^2 \right) \log \Gamma^2_1.
		\end{equation}
		If species \textsf{A} is chemostatted, we recover the fact that $ \mathcal{A}^*$ always saturates at $\log ( \beta/\alpha )$. 
		If species \textsf{B} is chemostatted, $\mathcal{A}^*$ is constrained to be between two values, attained at the edges of the feasible zone ($\Gamma^1_2 = 1$ or $\beta/\alpha$): 
		\begin{equation}
			\alpha \log \left(\dfrac{\beta}{\alpha} \right) < \mathcal{A}^\ast < \beta \log \left(\dfrac{\beta}{\alpha}\right),
		\end{equation}
		which is in agreement with what we found in Eq.~\ref{SM-eq:Upper_bound_Type_I} when $\beta = 2$ and $\alpha = 1$. 
		Note that, when $\alpha > 1$, the lower bound on $\mathcal{A}^*$ is necessarily greater than $\log \beta/\alpha $, which is in accordance with our principle on \emph{seed-dependent} mode of production. 
		
		%
		\section*{
			Supplementary Note 12:
			Introductory example
		}
		
		
		As claimed in the main text, the introductory example can be solved explicitly in the case $\alpha=1, ~\beta=2$, 
		\begin{equation}
			\label{SM-eq:Ex_network}
			\begin{split}
				\ce{\textsf{F} + \textsf{A}& <-->[$k_{+1}$][$k_{-1}$]   \textsf{B}}\\
				\ce{\textsf{B} & <-->[$k_{+2}$][$k_{-2}$] 2 \textsf{A} + \textsf{W}. }
			\end{split} 
		\end{equation}
		When species \textsf{B} is chemostatted, the steady-state concentration of \textsf{A}, $a$, is the root of a second-order polynomial: 
		\begin{equation}
			\label{SM-eq:ss_example_B}
			\d_t a = 0 = 2 \, k_{+2} \, b - 2 \, k_{-2} \, w \, a^2
			+ k_{-1} \, b - k_{+1} \, f \, a.
		\end{equation}
		The latter is easily solved and leads to the following expression for the macroscopic current:
		\begin{equation}
			\label{eq:macr_J_type_I_B}
			\mathcal{J}^\ss   = k_{+2} ~ b  - 
			\dfrac{\left( k_{+1} \, f \right) ^2}{16 \, k_{-2}  \, w} \cdot 
			\left(~  1 - \sqrt{1 + \dfrac{8 \, k_{-2} (2 \, k_{+2} + k_{-1})}{k_{+1}^2}\, \dfrac{b \, w}{f^2} } ~ \right)^2.
		\end{equation} 
		$\Js$ shares the same sign as the chemical affinity and has an extremum when
		\begin{equation}
			\frac{b^\ast \, w}{f^2} = \frac{1}{2}\, \frac{k_{+1}^2 k_{+2}}{k_{-1}^2 k_{-2}} \, \dfrac{ k_{+2} + k_{-1}}{2\, k_{+2} + k_{-1}}. 
		\end{equation}
		At this point, the reaction quotient is given by 
		\begin{equation}
			Q^* = \frac{K}{2} \, \dfrac{ k_{+2} + k_{-1}}{2 \, k_{+2} + k_{-1}} \leq \dfrac{K}{2}.
		\end{equation}
		Thus, when $\sf{B}$ is chemostated the optimal affinity is  bounded by a simple rate-independent constant: 
		\begin{equation}
			\mathcal{A}^\ast \geq  \ln 2.
		\end{equation}
		
		When one makes no assumption on $\alpha, ~\beta$ (with $0< \alpha < \beta$), the steady-state \eqref{SM-eq:ss_example_B} becomes a higher-order polynomial. 
		Computing the roots of such polynomials is in general difficult. This illustrates the need of a more tractable procedure to extract the bound on the chemical affinity, which would not rely  on the explicit solutions of  the steady-state problem as explained in the main text. 
		
		\paragraph{Analogous system based on specific degradation}
		
		Chemostatting  one of core species is not the only way to maintain Eq.~\eqref{SM-eq:Ex_network} away from equilibrium.
		Another important alternative way to do so is to introduce a specific degradation term.  
		For example, if one considers the degradation of species \textsf{A}, the dynamics becomes (for $\alpha=1$ and $\beta=2$):  
		\begin{align}
			\d_t\, a  & = 2\, j_2 - j_1 - \kappa \, a \\
			\d_t\, b & = j_1 - j_2 \label{eq:dyn_Type_I_deg_A_eqn_B},
		\end{align}
		where $\kappa$ is a parameter controlling the rate of degradation and the elementary currents are defined as in Eqs.~(5)-(6) of the main text.
		As in the case where $\textsf{A}$ were chemostatted, the steady-state fluxes are $ j_1^\ss = j_2^\ss = \mathcal{J}^\ss$, with 
		\begin{equation}
			\mathcal{J}^\ss (\kappa)  = \frac{\kappa}{k_{-1}  k_{-2} ~ w}\,  \left[k_{+1} k_{+2} ~ f  - \kappa \, \left(k_{+2} + k_{-1} \right) \right].
		\end{equation}
		As explained in the main text, $\Js$ vanishes when the degradation goes beyond a threshold $\kappa > \kappa_c  = \linebreak k_{+1} k_{+2}~ f/\left( k_{+2}  + k_{-1} \right)$.
		Unlike the previous case, $\kappa_c$ is not related to the equilibrium state of the system without degradation but is the maximum degradation rate the system can sustain. 
		For values $\kappa \geq \kappa_c $ degradation overcomes the autocatalytic chemical reactions, and the system reaches its zero fixed point. 
		Finally we noticed that the global flux $\mathcal{J}^\ss$ has a maximum at $\kappa^\ast = \kappa_c/2 $ or, equivalently, $\mathcal{A}^* = \log 2$. \\
		Introducing a specific degradation on $\textsf{B}$ instead, one would recover that the optimum global flux is attained for values of the affinity $ \log 2 \leq \mathcal{A}^* \leq \log 4$. 
		
		\section*{
			Supplementary Note 13:
			Bounds for the five simplest autocatalytic motifs} 
		
		We summarize in Table~1 of the materials and methods of the main text the value of the lower bound on $\mathcal{A}^\ast$ for the five simplest autocatalytic motifs described in \cite{Blokhuis2020}. 
		Being 3$\times$3 (except for Type~I that is 2$\times$2), these networks have a lower bound given by the determinant condition \eqref{SM-eq:general_bound_3x3} with $\S_+ = \text{Id.}$
		We have checked numerically that \eqref{SM-eq:general_bound_3x3} is the tightest, except for Type~V. 
		In that case, \eqref{SM-eq:general_bound_3x3} predicts $(\log 5)/2$ as lower bound while a numerical simulation shows that the actual one is $\log 3$.
		
		The tightest bound for the Type~V motif can be obtained by using the relation provided by the determinant condition and by combining it with the structural constraints (Eq.~(27) of the main text).
		Let us assume without loss of generality due to the symmetry of the network that \textsf{A} is chemostatted: $\boldsymbol{g}_\textsf{A} = (0, 1/2, 1/2)^\top$. 
		As a consequence, $ F_{-1} = F_{+1} $ or, equivalently, $\mathcal{A}_1^\ast = 0$ and $2\, \mathcal{A}^\ast = \mathcal{A}_2^\ast + \mathcal{A}_3^\ast$.
		Now,
		\begin{equation}
			\mathbb{M} = \begin{pNiceMatrix}
				1 & -1 & -1 \\ -1 & \eA{2} & -1 \\ -1 & -1 & \eA{3}	\end{pNiceMatrix},
		\end{equation}
		and therefore $\det \left( \mathbb{M} \right) = 0 \Leftrightarrow \exp(2\, \mathcal{A}^\ast) = 3 + \exp \left( \mathcal{A}_1^\ast \right)  + \exp \left( \mathcal{A}_3^\ast \right) $. From the positivity of $\mathcal{A}_1^\ast$ and $\mathcal{A}_3^\ast$, one would predict a lower bound for $\mathcal{A}^\ast$
		equal to $(\log 5)/2$. 
		On the other hand, the structural constraints read: 
		\begin{equation}
			\begin{cases}
				F_{-1} = F_{+2} + F_{+3} \\
				F_{-2} = F_{+1} + F_{+3} \\ 
				F_{-3} = F_{+1} + F_{+2},
			\end{cases} 
		\end{equation}
		which imply $e^{\mathcal{A}_2^\ast}=F_{-2}/F_{+2}=(F_{+1}+F_{+3})/F_{+2}$, and thus $x=F_{+3}/F_{+2}=(e^{\mathcal{A}_2^\ast}-1)/2 \ge 0$. Then, the expression for $\exp \left( 2 \mathcal{A}^\ast \right) $ can be re-written as: 
		\begin{equation} \nonumber 
			e^{2\mathcal{A}^\ast}= 3 + \dfrac{F_{-2}}{F_{+2}} + \dfrac{F_{-3}}{F_{+3}} 
			= 5 + 2 \left( x + \dfrac{1}{x} \right).
		\end{equation}
		Finally, since $x + x^{-1} \geq 2$ for all $x \ge 0$, we recover that the tightest bound - as confirmed numerically - is : 
		\begin{equation}
			e^{\mathcal{A}^\ast} \geq 3. 
		\end{equation}
		
		We now come back to the specific case of Type~I networks. 
		As explained in the main text, chemostatting species \textsf{B} turns the equality, $\mathcal{A}^\ast = \log 2$ into an inequality $\mathcal{A}^\ast \geq \log 2$.
		However, Figure~1.C of the main text also shows that the affinity is upper-bounded by $\log 4$, which strongly constrains the location of the extremum of the global rate. 
		This unexpected double constraint can also be obtained using the determinant condition. 
		In that case, the condition on the determinant yields
		\begin{equation}
			\label{SM-eq:Determinant_Type_I_B}
			e^{\mathcal{A}_1^\ast + \mathcal{A}_2^\ast} = 2. 
		\end{equation}
		However, $\boldsymbol{g}_\textsf{B} = (2, 1)^\top$ also implies that $\mathcal{A}^\ast = 2\, \mathcal{A}_1^\ast + \mathcal{A}_2^\ast$, and  $\mathcal{A}_1^\ast,~\mathcal{A}_2^\ast  > 0 $.
		Thus, from \eqref{SM-eq:Determinant_Type_I_B}, $\exp \left( \mathcal{A}_1^\ast \right) = 2 \exp \left( - \mathcal{A}_2^\ast \right)  \leq 2$ and similarly for $\exp \left( \mathcal{A}_2^\ast \right)$.
		Hence we obtain: 
		\begin{align}
			1 \leq e^{\mathcal{A}_1^\ast} \leq 2, & \quad  & 1 \leq e^{\mathcal{A}_2^\ast} \leq 2. 
		\end{align}
		Multiplying \eqref{SM-eq:Determinant_Type_I_B} by $e^{\mathcal{A}_1^\ast}$ yields the behavior shown in Fig.~1.C of the main text: 
		\begin{equation}
			\label{SM-eq:Upper_bound_Type_I}
			2 \leq e^{\mathcal{A}^\ast} \leq 4. 
		\end{equation}
		
		\section*{
			Supplementary Note 14:
			The Hinshelwood autocatalytic cycle}
		
		We finish by analyzing the Hinshelwood autocatalytic cycle, which is a model, in which a set of $n$ enzymes catalyzes each other in a cyclical manner \cite{Hinshelwood1952}. For $n=2$, this model can be written as : 
		\begin{equation}
			\begin{split}
				& \ce{\textsf{F}_1  <-->[\textsf{E}_1] \textsf{E}_2}, \\
				& \ce{\textsf{F}_2  <-->[\textsf{E}_2] \textsf{E}_1},
			\end{split}
		\end{equation}
		where $\textsf{F}_1$ and $\textsf{F}_2$ are two substrates and $\textsf{E}_1$ and $\textsf{E}_2$ are the enzymes.
		To adapt this model to the framework of \cite{Blokhuis2020}, we split every reaction above into two consecutive reactions with intermediates $\textsf{F}_1\textsf{E}_1$ and $\textsf{F}_2\textsf{E}_2$. Then, the reactions become 
		\begin{equation}
			\begin{split}
				& \ce{\textsf{E}_1 + \textsf{F}_1 <-->\textsf{F}_1\textsf{E}_1 <--> \textsf{E}_1 + \textsf{E}_2}, \\
				& \ce{\textsf{E}_2 +\textsf{F}_2 <-->\textsf{F}_2\textsf{E}_2 <--> \textsf{E}_1 + \textsf{E}_2}.
			\end{split}
		\end{equation}
		If we now assume that the substrates are in excess and we use the notation $\textsf{A}=\textsf{E}_1, \textsf{B}=\textsf{F}_1\textsf{E}_1, \textsf{C}=\textsf{E}_2$ and $\textsf{D}=\textsf{F}_2\textsf{E}_2$,
		we obtain the following reaction scheme and stoichiometric matrix :  
		\begin{align}
			\begin{cases}
				\ce{\textsf{A}  <-->[\mathsfit{1}] \textsf{B}} \\
				\ce{\textsf{B}  <-->[\mathsfit{2}] \textsf{A + C}} \\
				\ce{\textsf{C}  <-->[\mathsfit{3}] \textsf{D}} \\ 
				\ce{\textsf{D}  <-->[\mathsfit{4}] \textsf{A + C}}
			\end{cases}, & \quad & 
			\S =  \begin{pNiceMatrix}[last-row, first-col]
				\textsf{A} & -1 & 1 & 0 & 1  \\ 
				\textsf{B} & 1 & -1 & 0 & 0 \\ 
				\textsf{C} & 0 & 1 & -1 & 1  \\ 
				\textsf{D} & 0 & 0 & 1 & -1 \\
				& \mathsfit{1} & \mathsfit{2} & \mathsfit{3} & \mathsfit{4} \\
			\end{pNiceMatrix}.
		\end{align}
		The matrix inverse is:
		\begin{align}
			\S^{-1} =  \begin{pNiceMatrix}[last-row, first-col]
				\mathsfit{1} & 0 & 1 & 1 & 1  \\ 
				\mathsfit{2} & 0 & 0 & 1 & 1 \\ 
				\mathsfit{3} & 1 & 1 & 0 & 1  \\ 
				\mathsfit{4} & 1 & 1 & 0 & 0 \\
				& \boldsymbol{g}_\textsf{A} & \boldsymbol{g}_\textsf{B} & \boldsymbol{g}_\textsf{C} & \boldsymbol{g}_\textsf{D} \\
			\end{pNiceMatrix},
		\end{align}
		and the structural constraints are:
		\begin{equation}
			\label{SM-eq:Structural_constraints_Hinshelwood}
			\begin{cases}
				F_{-1} = F_{+2} \\ 
				F_{-2} = F_{+1} + F_{+3} \\ 
				F_{-3} = F_{+4} \\ 
				F_{-4} = F_{+1} + F_{+3}
			\end{cases}
		\end{equation}
		Applying the determinant condition on Hinshelwood gives:
		\begin{equation}
			\label{SM-eq:Hinshelwood_determinant}
			e^{\mathcal{A}_1^\ast + \mathcal{A}_2^\ast + \mathcal{A}_3^\ast + \mathcal{A}_4^\ast } - e^{\mathcal{A}_1^\ast + \mathcal{A}_2^\ast } - e^{\mathcal{A}_3^\ast + \mathcal{A}_4^\ast } =0,
		\end{equation}
		which leads to $ e^{\mathcal{A}^\ast} \geq 2$. 
		This bound, while correct, can be improved.			
		If \textsf{B} or \textsf{D} is chemostatted, only the reaction consuming the chemostat ($\mathsfit{2}$ or $\mathsfit{3}$) is at equilibrium. 
		As in the case of Type~V, the actual value of the bound is obtained using  the structural constraints in the relation on $\exp \left( \mathcal{A}^\ast \right)$ obtained from the determinant condition: 
		\begin{equation}
			e^{\mathcal{A}^\ast } = \dfrac{F_{-1} F_{-2}}{F_{+1}F_{+2}} + \dfrac{F_{-3} F_{-4}}{F_{+3}F_{+4}}.
		\end{equation}
		Incorporating the constraints \eqref{SM-eq:Structural_constraints_Hinshelwood} in this expression gives 
		\begin{equation}
			e^{\mathcal{A}^\ast} = 2 + \dfrac{F_{+3}}{F_{+1}} + \dfrac{F_{+1}}{F_{+3}} \geq 4,
		\end{equation}
		which is the tightest bound that can be obtained by varying the rate constants as shown on Figure~3.D in the main text.

		The situation is drastically different if species \textsf{A} or \textsf{C} are chemostatted.
		In that case, two reactions are at equilibrium (either $\mathsfit{1},\mathsfit{2}$ or $\mathsfit{3},\mathsfit{4}$),  implying that one of the optimal elementary affinities is diverging. 
		Indeed, assuming \textsf{A} is chemostatted, we have $g_\textsf{A}^\mathsfit{1} = g_\textsf{A}^\mathsfit{2} = 0$. This leads to two additional constraints:  $F_{-1} = F_{+1}$ and $F_{-2} = F_{+2}$.  Thus, 
		\begin{equation}
			\label{SM-eq:Hinshelwood_incompatible constraints}
			\begin{cases}
				F_{-1} = F_{+1} = F_{-2} = F_{+2} \\[0.8em] 
				F_{-2} = F_{+1} + F_{+3} 
			\end{cases},
		\end{equation} 
		which implies $F_{+3} = 0$.
		On the other hand, $\exp (\mathcal{A}_3^\ast) = F_{-3}/F_{+3}$, obtaining the divergence of $\exp (\mathcal{A}_3^\ast)$.
		The structural constraints cannot be satisfied and, as a consequence there is no extremum of the global flux within $]0, +\infty[$. 
		$\Js (Q)$ is a decreasing function of $Q$ being maximum at $Q^\ast = 0$. 
		Performing similar reasoning, one shows that controlling \textsf{C} instead leads to a divergence of $\exp (\mathcal{A}_1^\ast)$.

		\section*{
			Supplementary Note 15:
			A bi-stable network: the Schl\"{o}gl model} 
		
		We now study the robustness of our approach when  bistability is present in the dynamical system.  
		To illustrate this case we analyze the Schl\"{o}gl model:
		\begin{align}
			\begin{cases}
				\ce{\textsf{A}  <-->[\mathsfit{1}]  \textsf{X}} \\
				\ce{\textsf{2~X}  <-->[\mathsfit{2}]  \textsf{3~X}} \\
			\end{cases}, & \quad & \S_+ = ~ \begin{pNiceMatrix}[last-row, first-col]
				\textsf{A} & 1 & 0  \\ 
				\textsf{X} & 0 & 2 \\ 
				& \mathsfit{1} & \mathsfit{2} \\
			\end{pNiceMatrix}, & \quad & \S_- = ~ \begin{pNiceMatrix}[last-row, first-col]
				\textsf{A} & 0 & 0  \\ 
				\textsf{X} & 1 & 3 \\ 
				& \mathsfit{1} & \mathsfit{2} \\
			\end{pNiceMatrix},
			& \quad & \S^{-1} = ~ \begin{pNiceMatrix}[last-row, first-col]
				\mathsfit{1} & -1 & 0 \\ 
				\mathsfit{2} & 1 & 1 \\ 
				& \boldsymbol{g}_\textsf{A} & \boldsymbol{g}_\textsf{X} \\
			\end{pNiceMatrix}.
		\end{align}
		\begin{figure}[h]
			\includegraphics{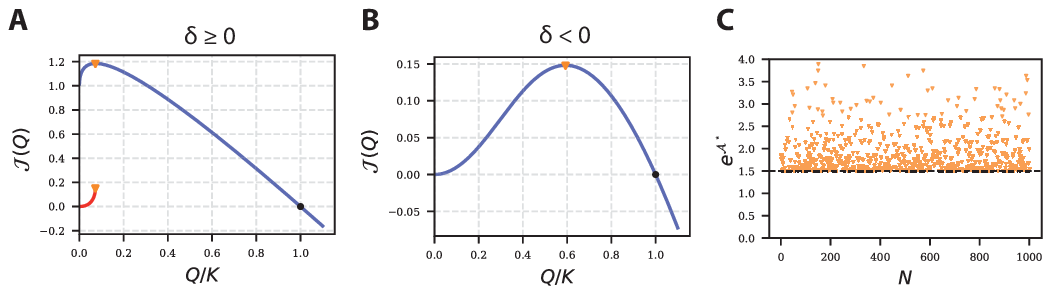}
			\caption{\label{fig:Schlogl}
				The location of the extremal flux of the Schl\"{o}gl cycle also exhibits a lower bound. 
				\textbf{A} When the second-order discriminant $\delta$ is positive, the third order discriminant \eqref{SM-eq:Third_order_discr} will have a positive region in which two values of the macroscopic flux co-exist (these curves were obtained for $k_{-1}=k_{-2}=1, ~ k_{+1}=0.1, ~ k_{+2}=2$).
				\textbf{B} If $\delta$ is negative, \eqref{SM-eq:Schlogl_cubic_eqn} has only one real root yielding a unique solution for the macroscopic flux (the curve was obtained for $k_{+1}=k_{+2} =k_{-2}= 1, ~ k_{-1}=2$). 
				In both cases the macroscopic fluxes vanish at equilibrium (solid black circle) and have a maximum (solid black square). 
				\textbf{C} The affinity at the maximum of the flux for randomly chosen kinetic constants (solid orange marker) are lower-bounded by 3/2 (dotted black curve), as predicted by \eqref{SM-eq:Schlogl_lower_bound}. 
			}
		\end{figure}
		Usually a second reactant \textsf{B} is found in the second reaction $\ce{2\textsf{X} + \textsf{B} <-->[\mathsfit{2}]  3\textsf{X}}$. 
		The latter is an \emph{external species}, whose concentration is absorbed it in the effective rate $k_{+2}$ and we do not note it explicitly for this reason.

		It is well-known that controlling the concentration of \textsf{A} in such network can result in bistable solutions for the stationary concentration of \textsf{X}, $x_\ss$.
		Indeed the latter obeys the following cubic equation:
		\begin{equation}
			\label{SM-eq:Schlogl_cubic_eqn}
			- k_{-2} ~ {x_\ss}^3 + k_{+2} ~ {x_\ss}^2 - k_{-1} ~ x_\ss + k_{+1} \, Q = 0,
		\end{equation}
		where $Q$ is the concentration of \textsf{A} that is also the reaction quotient.
		The cubic discriminant of \eqref{SM-eq:Schlogl_cubic_eqn} is: 
		\begin{equation}
			\label{SM-eq:Third_order_discr}
			\Delta = -27 \, k_{+1}^2\,  k_{-2}^2 ~ Q^2 + 2\, k_{+1}\,  k_{+2} \, \left( 9 \, k_{-1} \, k_{-2} - 2\,  k_{+2} \right) ~ Q + k_{+2}^2 \, k_{-1}^2 - 4 \, k_{-1}^3 \,  k_{-2}.
		\end{equation}
		If $\Delta \geq 0$, \eqref{SM-eq:Schlogl_cubic_eqn} has 3 real  solutions, and if $\Delta < 0 $  \eqref{SM-eq:Schlogl_cubic_eqn} has only 1 real  solution. 
		The third order discriminant \eqref{SM-eq:Third_order_discr} is a second order polynomial in $Q$. 
		Its sign depends on the sign of the second order discriminant $\delta = 16 \, k_{+1}^2 \, (k_{+2}^2 - 3\,  k_{-1}\,  k_{-2})^3$. 
		If $\delta > 0$ (resp. $\delta < 0$) then \eqref{SM-eq:Third_order_discr} will be positive (resp. negative) if $Q$ lies between (resp. outside) the two real roots of \eqref{SM-eq:Third_order_discr}. 
		In the bistable region where $\Delta > 0$, \eqref{SM-eq:Schlogl_cubic_eqn} two of the solutions to the third-order polynomial correspond to stable states, and the third one is an unstable state. Consequently, there will be  two different stable  fluxes in this parametric region (see Figure~\ref{fig:Schlogl}). 
		
		The matrix $\mathbb{M}$ is here given by 
		\begin{equation}
			\mathbb{M} = \begin{pNiceMatrix}
				\eA{1} & 0 \\ 0 &  \eA{2}
			\end{pNiceMatrix} - \left( \S_+^{-1} \cdot \S_- \right) = \begin{pNiceMatrix}  \eA{1} & 0 \\ -1/2 &  \eA{2} -  3/2 \end{pNiceMatrix}.
		\end{equation}
		The condition $\det \left( \mathbb{M} \right) = 0$ yields $\exp \left( \mathcal{A}_1^\ast + \mathcal{A}_2^\ast \right) = (3/2) \exp(\mathcal{A}_1^\ast) $. 
		In addition, because  $\boldsymbol{g}_\textsf{A} = (-1, 1)^\top$, the cycle affinity at steady-state is $\mathcal{A}^\ast = \mathcal{A}_2^\ast - \mathcal{A}_1^\ast$. This leads to  the following lower bound: 
		\begin{equation}
			\label{SM-eq:Schlogl_lower_bound}
			e^{\mathcal{A}^\ast} = \dfrac{3}{2} e^{-\mathcal{A}_1^\ast} \geq \dfrac{3}{2}.
		\end{equation}
		Our formalism applies to situations where the derivative of the flux is zero. This can happen only for the branch associated to the solution of \eqref{SM-eq:Schlogl_cubic_eqn} which has the highest rate.  The lower flux is an ever-increasing function of $Q$  (see Fig.~\ref{fig:Schlogl}A).
		As a consequence, it will reach a maximum at the transition between the bistable region and the monostable region, where the corresponding steady-state disappears.

		\end{document}